\def\arxivbuild{}
\documentclass[12pt]{article}
\ifdefined\pdfminorversion\pdfminorversion=7\fi

\usepackage[T1]{fontenc}
\usepackage[utf8]{inputenc}
\usepackage{lmodern}
\usepackage[a4paper,margin=1in]{geometry}
\usepackage{amsmath}
\usepackage{amssymb}
\usepackage{bm}
\usepackage{graphicx}
\graphicspath{{figures/}}
\usepackage{booktabs}
\usepackage{multirow}
\usepackage{array}
\usepackage{makecell}
\usepackage{threeparttable}
\usepackage[numbers,sort&compress]{natbib}
\usepackage{lineno}
\usepackage{microtype}
\ifdefined\arxivbuild
  \usepackage{docmute,float}
  \newcommand{\SIref}[1]{\ref{#1}}
\else
  \usepackage{xr-hyper}
  \newcommand{\SIref}[1]{\ref{si-#1}}
\fi
\usepackage[hidelinks]{hyperref}
\ifdefined\arxivbuild\else
\fi
\usepackage{placeins}


\title{Retaining intermediate element identities enables longer training to improve crystal discovery}
\author{%
Yucheng Jin\textsuperscript{1}, Chengqian Zhang\textsuperscript{1,2}, Han Wang\textsuperscript{3,4,*}\\[8pt]
{\small
\begin{minipage}{0.92\textwidth}\centering
\textsuperscript{1}AI for Science Institute, Beijing, China\\
\textsuperscript{2}Center for Data Science, Peking University, Beijing, China\\
\textsuperscript{3}National Key Laboratory of Computational Physics, Institute of Applied Physics and Computational Mathematics, Beijing, China\\
\textsuperscript{4}HEDPS, CAPT, College of Engineering, Peking University, Beijing, China\\[4pt]
\end{minipage}}
}
\date{}

\begin{document}
\maketitle
\ifdefined\arxivbuild\else
\linenumbers
\fi

\setcounter{secnumdepth}{0}

\begin{abstract}
Longer training of crystal diffusion models can improve stability while reducing novelty, limiting further gains in the fraction of metastable, unique and novel (MSUN) candidates.
Here we introduce Intermediate Identity-Locking Sampling (IILS), which fixes identities projected from noisy intermediate element states while continuing to denoise atomic coordinates and the lattice, without modifying the trained model.
With a chemically informed element representation, IILS maintains high novelty and enables further MSUN gains after performance under full sampling has largely saturated.
It achieves 53.0\% MSUN on MP20, compared with 37.1\% for the best reported model in our benchmark comparison.
Its MSUN advantage extends across the tested architectures and datasets.
These findings demonstrate that saturation of MSUN under full sampling does not preclude further improvements in crystal discovery through longer training.
\end{abstract}

Generative models offer a route to exploring the vast space of inorganic crystals by learning compositional and structural regularities from known materials~\cite{nouira2019crystalgan,noh2019imatgen,zhao2021cubicgan,ren2022ftcp,xie2022cdvae,zhao2023pgcgm}.
Recent approaches include autoregressive models of crystal representations~\cite{xiao2023slices,antunes2024crystallm,gruver2024crystallm,choudhary2024atomgpt,cao2025crystalformer,kazeev2025wyformer,xu2025plaidpp}, diffusion models~\cite{jiao2023diffcsp,yang2024unimat,jiao2024diffcsppp,zeni2025mattergen,levy2025symmcd,park2025chemeleon,joshi2025adit,okhotin2025miad,yi2026crystaldit,veljkovic2026crystalite}, and methods based on flow matching and stochastic interpolants~\cite{miller2024flowmm,sriram2024flowllm,luo2025crystalflow,hollmer2025omatg}.
Hybrid pipelines also combine composition generation with structural diffusion~\cite{cheng2026crysvcd}.
For computational materials discovery, candidates should satisfy energetic stability criteria, be novel relative to reference structures and avoid duplication within the generated set.
Improvements in validity or stability alone do not necessarily increase the fraction satisfying these requirements jointly.
Recent studies assess this joint success through the metastable, unique and novel (MSUN) fraction~\cite{park2026chemeleon2,betala2025lemat}.
Increasing this fraction is therefore a central objective of generative crystal discovery.

Longer training can reduce test loss in language models~\cite{kaplan2020scaling} and improve sample quality in image diffusion models~\cite{peebles2023dit}.
For crystal discovery, however, longer training need not improve the MSUN fraction.
Standard diffusion training optimizes denoising on known crystal structures, without directly maximizing joint discovery metrics~\cite{jiao2023diffcsp,zeni2025mattergen}.
For Crystalite, training can increase the fraction of candidates meeting the metastability criterion while reducing uniqueness and novelty.
Under one loss-weight setting, the MSUN fraction peaks and then declines~\cite{veljkovic2026crystalite}.
CrystalDiT likewise reports improving structural and compositional validity alongside a declining unique-and-novel fraction during training~\cite{yi2026crystaldit}.
Such trends limit the extent to which further training translates into gains in materials discovery.

Several strategies have been proposed to address this trade-off.
Selecting checkpoints using discovery-related metrics balances quality and novelty across training stages, but does not improve the sampling performance of later checkpoints~\cite{yi2026crystaldit}.
Reducing the weight of element denoising relative to geometry losses can sustain MSUN, but requires adjusting the training objective~\cite{veljkovic2026crystalite}.
Reinforcement-learning post-training can increase MSUN through rewards for stability, novelty and diversity, at the cost of additional candidate generation, reward evaluation and parameter updates~\cite{park2026chemeleon2,prakash2026omatgrpo}.
We investigate whether a model trained with the original denoising objective can achieve higher MSUN through more effective use of its generation process, without additional parameter optimization.

Here we demonstrate that, with a chemically informed element representation, longer training can increase MSUN when intermediate element identities are retained, even after MSUN under full sampling has largely saturated.
To use these intermediate element identities for crystal discovery, we introduce Intermediate Identity-Locking Sampling (IILS), which fixes identities projected from noisy intermediate states while continuing to denoise atomic coordinates and the lattice, without updating the model parameters.
The benefits depend on the element representation and projection time.
We introduce Chemical Similarity-based Spherical Encoding (CSSE) to encode chemical relationships learned by pretrained models.
Across the tested architectures and datasets, CSSE with IILS yields higher MSUN than the corresponding full-sampling configurations.
In property-guided generation, IILS also increases the fraction of candidates meeting both discovery and property criteria for the tested band gap and magnetic density targets.

\section{Results}

\subsection{Intermediate element identities enable discovery gains from longer training}

\begin{figure}[!tp]
    \centering
    \includegraphics[width=\textwidth]{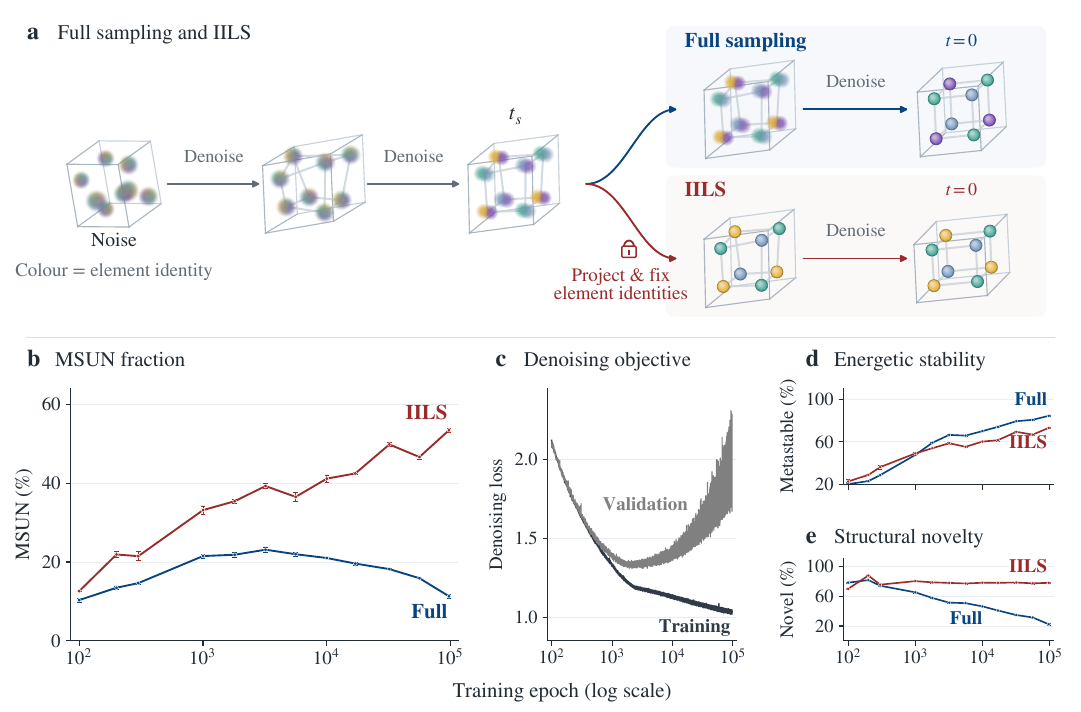}
    \caption[MSUN during training under Full and IILS]{%
    \textbf{MSUN during training under full sampling and IILS.}
    \textbf{a,} Full sampling (Full) and Intermediate Identity-Locking Sampling (IILS).
    IILS fixes element identities at $t_s$ while coordinates and lattice continue denoising.
    Diffuse and sharp colours denote continuous element states and discrete identities.
    \textbf{b,} MSUN for DiffCSP with CSSE-CSP on MP20.
    \textbf{c,} Training and validation losses for the original denoising objective.
    \textbf{d,e,} Metastable and Novel fractions.
    Logarithmic epoch axes.
    Panels \textbf{b,d,e} show twelve checkpoints between epochs 100 and 100,000; each point is the mean over three independent training runs and the error bars are standard errors.
    Full uses $\sigma=1$, and IILS uses $\sigma=0.2$ and $t_s=0.2$.
    MSUN denotes the fraction simultaneously satisfying metastability, uniqueness and novelty, with $E_{\mathrm{hull}}\leq0.1\,\mathrm{eV\,atom^{-1}}$ for metastability.
    }
    \label{fig:training-discovery}
\end{figure}

Retaining intermediate element identities enables further MSUN gains from longer training after performance under full sampling has largely saturated.
For the diffusion models studied here, full sampling (Full) jointly denoises continuous element states, atomic coordinates and the lattice to completion~\cite{jiao2023diffcsp,veljkovic2026crystalite}.
IILS follows the same trajectory up to an intermediate time $t_s$, projects the element states to discrete identities and fixes those identities while coordinates and lattice continue to denoise (Fig.~\ref{fig:training-discovery}a).
The projection changes neither coordinates nor lattice, requires no additional parameter optimization, and incurs little additional sampling cost compared with Full (Supplementary Note~\SIref{sec:supp_sampling_efficiency}).

For DiffCSP, a diffusion-based crystal generator~\cite{jiao2023diffcsp}, MSUN increases under both Full and IILS during early training (Fig.~\ref{fig:training-discovery}b).
The model is trained on MP20, a collection of inorganic crystals with at most 20 atoms per unit cell drawn from the Materials Project database~\cite{jain2013materialsproject}.
It uses our CSSE-CSP element representation, which encodes relationships learned by a pretrained crystal structure prediction model (Methods and Fig.~\ref{fig:representation-timing}).
With further training, MSUN under Full plateaus over the first 10,000 epochs and then declines, whereas IILS continues to gain.
Between epochs 1,000 and 100,000, MSUN changes from 21.5\% to 11.3\% under Full and increases from 33.2\% to 53.0\% under IILS.
Training comparisons for Crystalite on MP20 and DiffCSP on AlexMP20 are reported in Supplementary Note~\SIref{sec:supp_continued_training}.
Following an initial decrease, validation loss for the original denoising objective rises while training loss continues to decrease (Fig.~\ref{fig:training-discovery}c).
This upturn approximately coincides with MSUN saturation under Full, whereas IILS yields further MSUN gains over the same training interval (Fig.~\ref{fig:training-discovery}b).

Both samplers produce an increasing fraction of candidates meeting the metastability criterion $E_{\mathrm{hull}}\leq0.1\,\mathrm{eV\,atom^{-1}}$ (Fig.~\ref{fig:training-discovery}d).
Under Full, this improvement is accompanied by declining novelty and, later in training, by declining MSUN.
Under IILS, novelty remains near 80\% during later training, while the metastable and MSUN fractions increase further (Fig.~\ref{fig:training-discovery}e).

\begin{figure}[!tp]
    \centering
    \includegraphics[width=\textwidth]{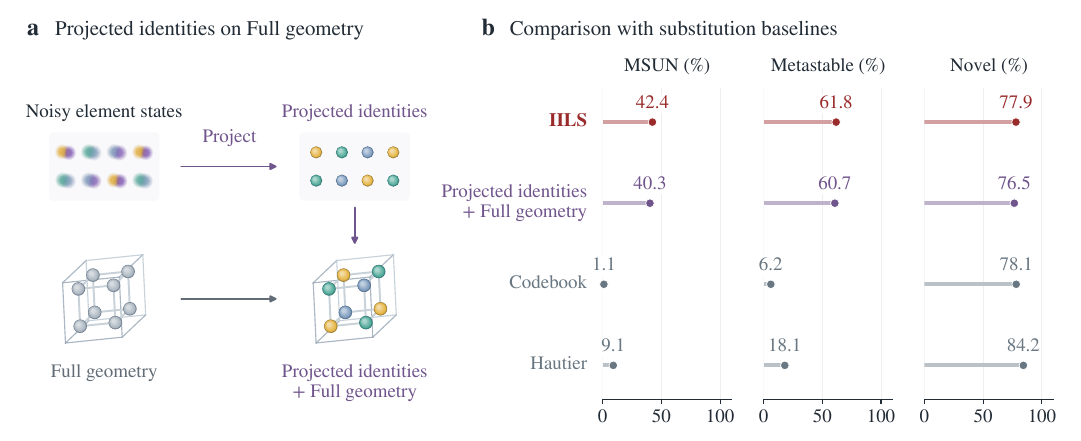}
    \caption[Intermediate element identities on Full geometries]{%
    \textbf{Useful element identities from intermediate states.}
Results for DiffCSP with CSSE-CSP on MP20 ($t_s=0.2$).
\textbf{a,} Noisy element states are projected to their nearest codewords.
The resulting identities are combined with the same Full trajectory's terminal coordinates and lattice (Projected identities + Full geometry).
\textbf{b,} Relaxed-crystal metrics for IILS and terminal-replacement controls.
Codebook and Hautier controls select sites randomly and replacements from empirical distributions.
    }
    \label{fig:intermediate-assignments}
\end{figure}

The training comparison establishes the MSUN benefit of IILS but does not separate the contribution of retained element identities from that of subsequent geometric denoising.
To examine the contribution of retaining these identities, we project the noisy element states at $t_s=0.2$ to their nearest element codewords and assign the resulting identities to the terminal coordinates and lattice of the corresponding Full trajectory (Fig.~\ref{fig:intermediate-assignments}a).
This construction, termed Projected identities + Full geometry, shares its geometry with the Full output before relaxation.
After relaxation, it achieves 40.3\% MSUN, compared with 42.4\% for IILS (Fig.~\ref{fig:intermediate-assignments}b).
This comparison indicates that retaining intermediate element identities is a key contributor to the high MSUN under IILS.
The similar performance on Full geometries suggests that denoising coordinates and lattice under fixed identities provides a modest additional benefit in this setting.

We further examine how intermediate element assignments evolve with training (Supplementary Fig.~\SIref{fig:supp_assignments_training}).
As training progresses, fewer atomic sites differ in identity from their full-sampling counterparts, while the remaining changes become increasingly enriched in empirically compatible substitutions. 
Meanwhile, the metastable and MSUN fractions of Projected identities + Full geometry also increase progressively with training. 
These results suggest that continued training produces more effective intermediate element assignments, which may help explain the sustained MSUN gains under IILS.

We then test whether static element-substitution rules can reproduce this performance on Full terminal geometries.
We replace identities at randomly selected sites, drawing replacement elements according to either similarities between CSSE-CSP element codewords (Codebook) or empirical ionic-substitution statistics (Hautier)~\cite{hautier2011substitution}.
Both controls target a substitution fraction of 50\%, close to the 48.8\% of sites whose element identities differ between paired IILS and Full outputs.
These controls yield only 1.1\% and 9.1\% MSUN, respectively, despite high novel fractions.
Their metastable fractions are 6.2\% and 18.1\%, compared with 60.7\% for Projected identities + Full geometry (Fig.~\ref{fig:intermediate-assignments}b).
Re-noising the coordinates and lattice to $t=0.2$ and denoising them with the substituted identities fixed yields only 1.9\% and 9.4\% MSUN after relaxation.
The tested substitution rules therefore do not recover the high MSUN obtained with intermediate element identities, even when followed by geometric denoising.

We also vary the fraction of randomly selected sites whose element identities are projected and fixed at $t_s=0.2$, while the remaining element states continue to denoise together with coordinates and lattice.
Increasing this fraction from 0.1 to 1.0 generally increases novelty and MSUN while decreasing the metastable fraction (Supplementary Fig.~\SIref{fig:supp_partial_projection}).

\FloatBarrier

\subsection{Element representation and timing shape IILS performance}

The preceding comparisons show MSUN gains from retaining intermediate element identities.
Because IILS selects the nearest element codeword at each site, we next examine whether these gains depend on the chemical relationships encoded by the codebook.
Our comparison includes chemically informed element representations used in CrystalDiT and Crystalite~\cite{yi2026crystaldit,veljkovic2026crystalite}, our Chemical Similarity-based Spherical Encoding (CSSE), and one-hot and randomized controls.

CSSE encodes element relationships learned through pretraining on other atomistic tasks as a chemical prior for crystal diffusion.
CSSE-CSP is derived from a crystal structure prediction model~\cite{jiao2023diffcsp}, and CSSE-MLIP from an interatomic potential trained to predict energies and forces~\cite{zhang2026dpa3}.
The design aims to retain the relational structure of the learned element representations while controlling their dimensionality and norm.
Pairwise cosine similarities characterize these relationships independently of the magnitudes of the source vectors.
A low-dimensional spherical codebook is constructed to approximate this similarity structure, compressing the relational information and giving all elements the same codeword norm.
Element relationships are thus encoded through angular separation, and the codebook remains fixed during diffusion-model training (Methods).

\begin{figure}[!tp]
    \centering
    \includegraphics[width=\textwidth]{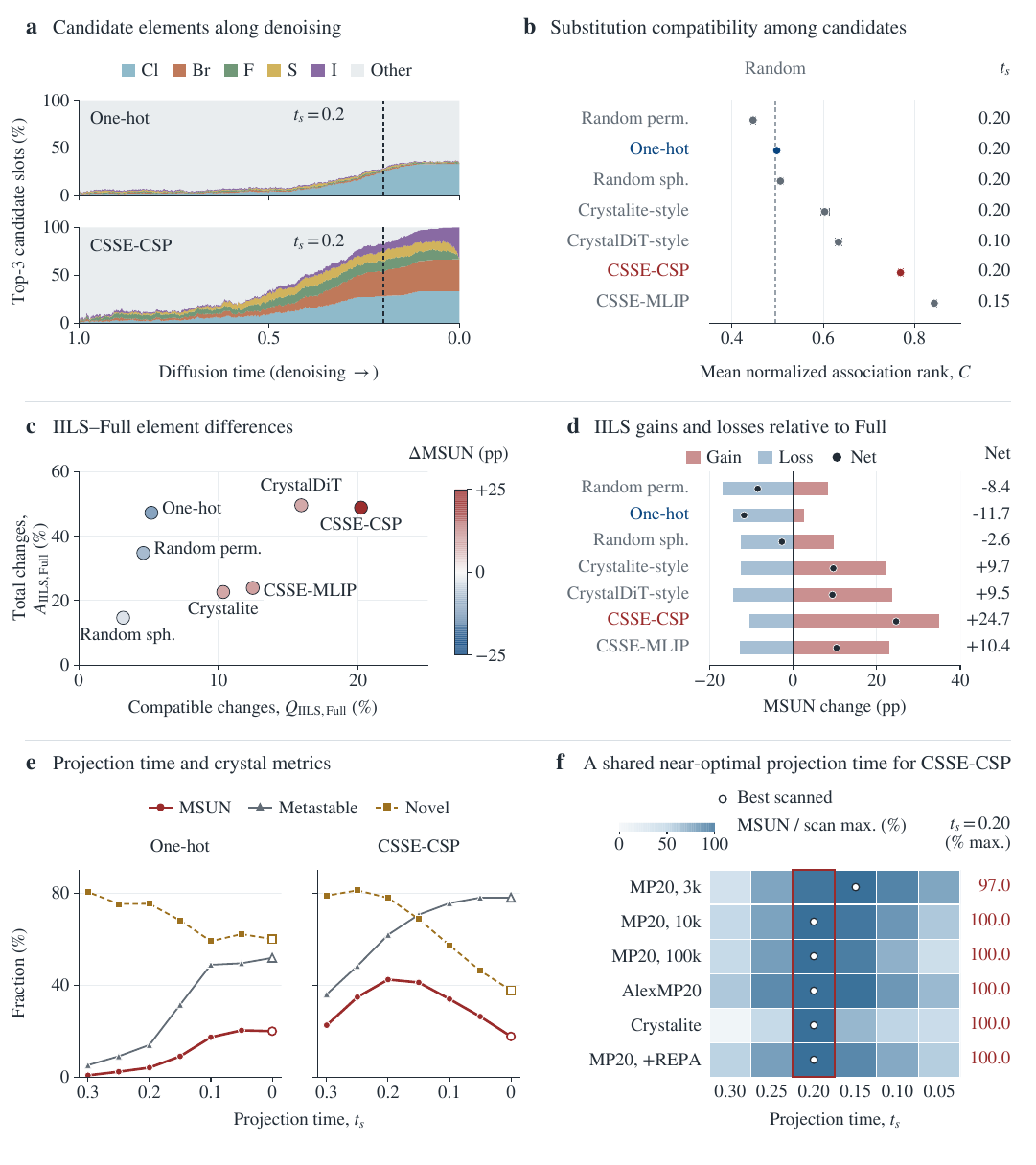}
    \caption[Element representation and projection time]{%
    \textbf{Element representation and projection time shape IILS performance.}
    Panels \textbf{a--e} use DiffCSP on MP20.
\textbf{a,} Top-three candidate-slot frequencies at sites ending as Cl under Full.
Dashed lines mark $t_s=0.2$.
\textbf{b,} Mean normalized association rank $C$, with 95\% confidence intervals over sites.
The dashed line denotes random triplets.
\textbf{c,} Total identity changes $A_{\mathrm{IILS,Full}}$ versus compatible changes $Q_{\mathrm{IILS,Full}}$, both expressed as fractions of all sites.
Colour denotes the MSUN difference between IILS and Full (IILS minus Full, in percentage points).
\textbf{d,} Fractions of paired outputs meeting the MSUN criteria only under IILS (gain) or only under Full (loss).
Dots show gain minus loss in percentage points.
Panels \textbf{c} and \textbf{d} summarize site-level identity changes and crystal-level MSUN gains and losses, respectively, for the same paired IILS and Full outputs.
\textbf{e,} Crystal metrics versus projection time, approaching Full at $t_s=0$.
\textbf{f,} CSSE-CSP MSUN across training lengths, datasets, architectures and an auxiliary REPA objective, normalized to each scan's maximum.
Circles mark maxima.
The red box and right column show $t_s=0.2$ and its normalized MSUN.
    }
    \label{fig:representation-timing}
\end{figure}

We first examine whether the learned relationships are reflected in the candidate elements near noisy intermediate states.
We identify these candidates by ranking element codewords by their distance from the state.
For sites whose Full terminal identity is chlorine (Cl), the three nearest codewords at $t=0.2$ more frequently include the other halogens Br, F and I with CSSE-CSP than with one-hot encoding (Fig.~\ref{fig:representation-timing}a).
The top-three statistics describe the candidate neighbourhood, whereas IILS projects each state to its single nearest codeword.

This organisation extends beyond sites ending as Cl.
We quantify it using the mean normalized association rank $C$, which averages normalized ranks of empirical substitution associations~\cite{hautier2011substitution} over the three candidate pairs at each site and then over sites (Methods).
Higher values indicate stronger empirical substitution associations among the candidate elements.
At the projection times indicated in Fig.~\ref{fig:representation-timing}b, $C$ is 0.769 for CSSE-CSP and 0.843 for CSSE-MLIP.
Both exceed the one-hot value of 0.498 and the uniformly random triplet reference of 0.495.
The Crystalite-style and CrystalDiT-style representations also exceed this reference.
The random-permutation control, which reassigns element labels while preserving the CSSE-MLIP codeword set, gives $C=0.446$.

We next compare paired IILS and Full outputs across codebooks to assess the extent and chemical compatibility of differences in element identities (Fig.~\ref{fig:representation-timing}c).
We denote the fraction of all atomic sites with different final element identities by $A_{\mathrm{IILS,Full}}$.
The fraction of all sites that both differ and satisfy the empirical compatibility criterion is $Q_{\mathrm{IILS,Full}}$.
This criterion requires the two elements to share a common oxidation state and meet the empirical substitution-association threshold (Methods).
Colours indicate $\Delta\mathrm{MSUN}$ in percentage points, defined for each codebook as the MSUN fraction under IILS minus that under paired Full sampling of the same trained model.

CSSE-CSP, CrystalDiT-style and one-hot change identities at similar fractions of sites (47--50\%), but compatible changes account for 20.2\%, 15.9\% and 5.2\% of all sites, respectively.
The first two representations increase MSUN, whereas one-hot decreases it.
Crystalite-style and CSSE-MLIP also increase MSUN while changing identities at about 23--24\% of sites, with compatible changes accounting for 10.3\% and 12.5\% of all sites, respectively.
Random spherical codewords and randomly permuted element labels give lower compatible-change fractions (3.2\% and 4.6\%) and negative MSUN changes.
The random-permutation control retains the CSSE-MLIP codeword set, yet produces more total identity changes and fewer compatible changes than CSSE-MLIP, accompanied by a reversal in the sign of the MSUN change.
Across these comparisons, positive MSUN changes are associated with a larger fraction of compatible identity changes, whereas the total extent of identity changes alone does not distinguish beneficial from detrimental interventions.

Chemically informed codebooks yield a larger fraction of MSUN gains under IILS than one-hot and randomized controls (Fig.~\ref{fig:representation-timing}d and Supplementary Note~\SIref{sec:supp_compatibility_gains}).
Gains occur when an IILS output meets the MSUN criteria and its corresponding Full output does not.
Losses occur when a Full output meets the criteria and its corresponding IILS output does not.
Gains range from 22.1\% to 35.0\% for Crystalite-style, CrystalDiT-style and the two CSSE codebooks, compared with 2.7--9.9\% for one-hot and the two random controls.
The corresponding loss fractions span overlapping ranges of 10.3--14.4\% and 12.5--16.8\%, respectively.
For CSSE-CSP, gains of 35.00\% outweigh losses of 10.33\%, giving a net increase of 24.67 percentage points.
Supplementary Note~\SIref{sec:supp_compatibility_gains} extends the $A$, $Q$ and $Q/A$ analysis and the corresponding gain--loss decomposition across the projection-time scan.

The MSUN benefit of IILS also depends on when element identities are fixed (Fig.~\ref{fig:representation-timing}e).
With CSSE-CSP, $t_s=0.2$ gives the highest MSUN among the seven tested times, 42.4\%, compared with 17.6\% under Full ($t_s=0$).
At the same projection time, one-hot MSUN instead falls well below its own Full value.
The two representations are comparably novel at that time, whereas their metastable fractions differ sharply.
The higher MSUN with CSSE-CSP is thus associated with retaining a much larger metastable fraction at comparable novelty.

This contrast is consistent with the stronger substitution associations among CSSE-CSP candidates (Fig.~\ref{fig:representation-timing}a,b).
Encoding chemical similarity in codeword distances may allow intermediate projection to retain chemically compatible alternatives before denoising has resolved the element identities.
One-hot codewords are equidistant and encode no such preference.
Moving projection earlier from $t_s=0.2$ to $t_s=0.3$ increases novelty only slightly, from 77.9\% to 78.8\%, while the metastable fraction falls from 61.8\% to 36.0\% and MSUN from 42.4\% to 22.5\% (Fig.~\ref{fig:representation-timing}e).
With a chemically informed codebook, selecting the projection time therefore amounts to balancing novelty against metastability (Supplementary Note~\SIref{sec:supp_projection_time}).

With CSSE-CSP, a shared projection time remains near-optimal across the tested settings (Fig.~\ref{fig:representation-timing}f and Supplementary Note~\SIref{sec:supp_projection_time}).
These comparisons vary training length (3,000, 10,000 and 100,000 epochs), dataset and model architecture, and include training with an auxiliary representation-alignment objective (REPA, Supplementary Note~\SIref{sec:supp_repa}).
Among the scanned times, $t_s=0.2$ gives the highest MSUN in five of the six settings.
For DiffCSP trained on MP20 for 3,000 epochs, the maximum occurs at $t_s=0.15$, but $t_s=0.2$ retains approximately 97\% of that maximum.
These results indicate that a common projection time can be reused across the tested CSSE-CSP settings, reducing the need for separate tuning.

\subsection{Extending the novelty--metastability Pareto frontier}

\begin{figure}[!tp]
    \centering
    \includegraphics[width=\textwidth]{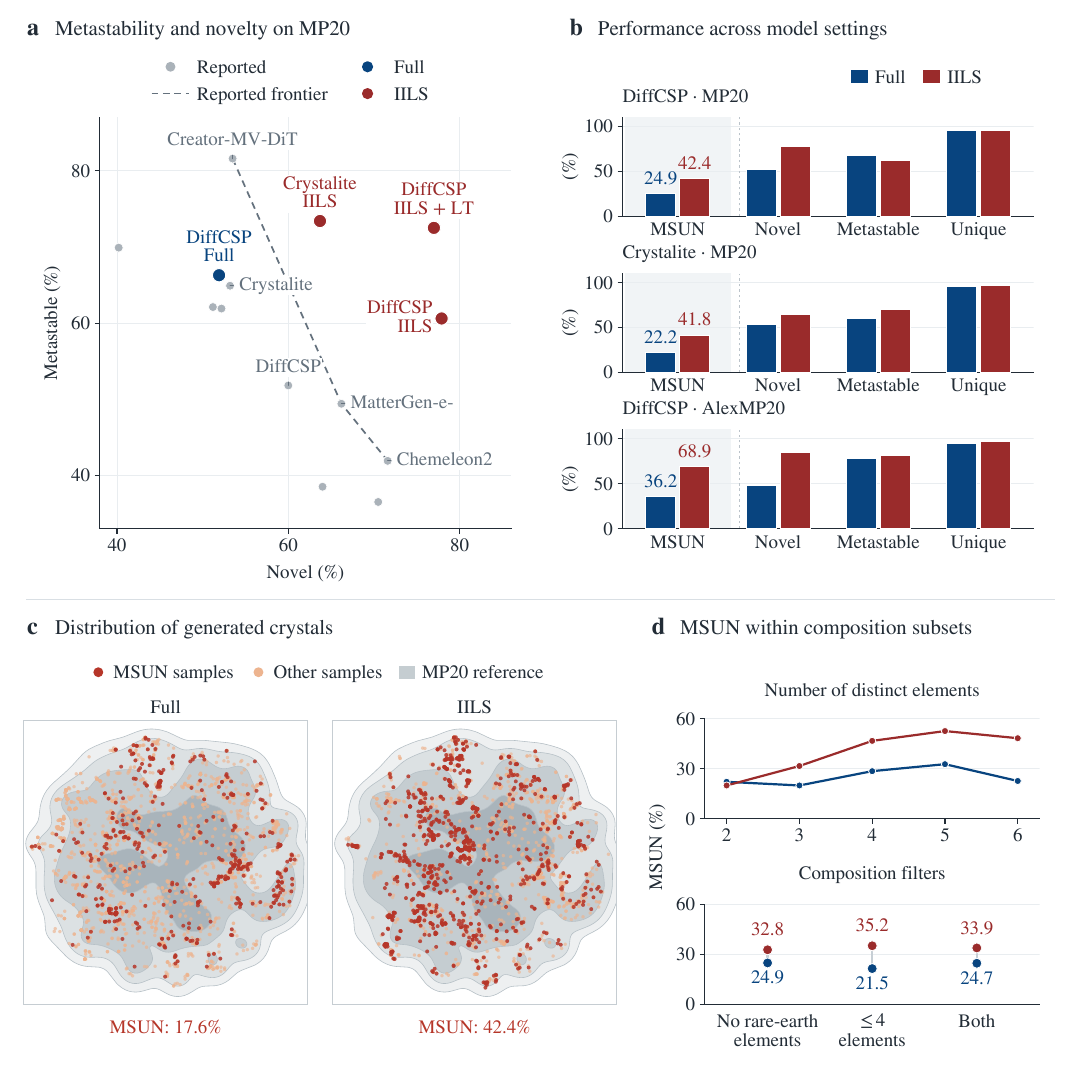}
    \caption[Discovery performance across models, datasets and compositions]{%
    \textbf{Discovery performance across models, datasets and compositions.}
\textbf{a,} Metastable and Novel fractions for ten reported MP20 benchmark configurations (grey) and this work's CSSE-CSP models (Full in blue and IILS in red).
The dashed line is the frontier of the reported leaderboard entries, computed on the full retained set before the view was cropped.
LT denotes longer training.
DiffCSP Full and IILS use different sampling noise scales.
\textbf{b,} MSUN, Novel, Metastable and Unique for three architecture--dataset settings with CSSE-CSP.
Values are means of three independent sampling runs.
Numbers label MSUN.
\textbf{c,} Joint t-SNE of DPA3 structure features for MP20 training data and one batch of 2,500 samples per method from the same CSSE-CSP DiffCSP checkpoint ($\sigma=0.2$).
Both views share axes and MP20 density contours.
Red marks samples meeting the MSUN criteria.
MSUN percentages are computed as in Table~\ref{tab:benchmark}, except the Full value, which is measured on the batch plotted here.
\textbf{d,} MSUN within groups containing two to six distinct elements (top) and after composition filtering (bottom), for DiffCSP with CSSE-CSP on MP20.
Filters exclude rare-earth-containing compositions, compositions with more than four elements, or both.
    }
    \label{fig:performance-transfer}
\end{figure}

With CSSE-CSP, IILS substantially extends the empirical Pareto frontier for novelty and metastability on MP20 (Fig.~\ref{fig:performance-transfer}a).
Both DiffCSP and Crystalite achieve higher novel and metastable fractions than their respective reported configurations in LeMat-GenBench~\cite{betala2025lemat}.
DiffCSP with IILS achieves the highest novel fraction among all plotted configurations and a metastable fraction more than 20 percentage points above that of Chemeleon2, the most novel benchmark configuration.
Crystalite with IILS reaches a higher metastable fraction at lower novelty than DiffCSP with IILS.
These IILS configurations lie well beyond the benchmark frontier, whereas the CSSE-CSP Full configuration of DiffCSP extends it only slightly.
For DiffCSP, longer training under the original denoising objective raises the metastable fraction while largely preserving novelty, further extending the frontier (DiffCSP IILS + LT in Fig.~\ref{fig:performance-transfer}a).

\begin{table}[p]
\centering
\begin{threeparttable}
\caption{\textbf{Crystal generation performance across models and sampling settings.}}
\label{tab:benchmark}
\footnotesize
\setlength{\tabcolsep}{3.0pt}
\renewcommand{\arraystretch}{1.12}
\begin{tabular*}{\textwidth}{@{}l c @{\extracolsep{\fill}} rrrrrrr@{}}
\toprule
Model / representation & Sampling & Valid$\uparrow$ & Unique$\uparrow$ & Novel$\uparrow$ & Stable$\uparrow$ & SUN$\uparrow$ & Metastable$\uparrow$ & MSUN$\uparrow$ \\
\midrule
\multicolumn{9}{@{}l}{\textbf{Reported methods | MP20}} \\
\addlinespace[2pt]
Creator-MV-DiT~\cite{creator2026creator} & --- & 99.6 & 97.8 & 53.5 & 6.2 & 2.8 & 81.6 & 37.1 \\
Crystalite~\cite{veljkovic2026crystalite} & --- & 97.2 & 95.8 & 53.2 & 5.7 & 1.2 & 64.9 & 24.8 \\
Chemeleon2~\cite{park2026chemeleon2} & --- & 95.2 & 88.1 & 71.6 & 0.0 & 0.0 & 41.9 & 22.3 \\
MatterGen~\cite{zeni2025mattergen} & --- & 95.7 & 95.1 & 70.5 & 0.5 & 0.1 & 36.5 & 15.9 \\
OMatG-FC~\cite{hollmer2025omatg} & --- & 97.2 & 92.8 & 28.9 & 11.7 & 1.4 & 77.4 & 14.6 \\
ADiT~\cite{joshi2025adit} & --- & 90.6 & 87.8 & 26.0 & 0.1 & 0.0 & 37.2 & 1.0 \\
PLaID++~\cite{xu2025plaidpp} & --- & 96.0 & 77.8 & 24.2 & 6.4 & 0.9 & 74.0 & 8.7 \\
MiAD~\cite{okhotin2025miad} & --- & 96.2 & 94.3 & 40.2 & 1.6 & 0.6 & 69.9 & 18.8 \\
\midrule
\multicolumn{9}{@{}l}{\textbf{Template-based baseline (this work) | MP20}} \\
\addlinespace[2pt]
Ion exchange~\cite{szymanski2025baselines} & --- & 99.9 & 99.9 & 56.8 & 5.7 & 2.2 & 67.1 & 31.9 \\
\midrule
\multicolumn{9}{@{}l}{\textbf{DiffCSP (this work) | MP20}} \\
\addlinespace[2pt]
One-hot & Full & 96.9 & 94.0 & 60.0 & 4.0 & 0.6 & 51.8 & 19.9 \\
 & IILS & 94.2 & 94.1 & 73.0 & 0.7 & 0.0 & 16.3 & 4.1 \\
\addlinespace[2pt]
Crystalite-style & Full & 97.6 & 95.7 & 50.9 & 4.5 & 0.8 & 66.6 & 22.3 \\
 & IILS & 97.6 & 92.2 & 64.8 & 3.4 & 0.9 & 57.2 & 27.8 \\
\addlinespace[2pt]
CrystalDiT-style & Full & 98.2 & 94.0 & 54.2 & 3.9 & 0.8 & 66.3 & 24.6 \\
 & IILS & 99.2 & 90.9 & 76.8 & 1.1 & 0.4 & 51.8 & 32.6 \\
\addlinespace[2pt]
CSSE-MLIP & Full & 97.2 & 95.7 & 54.9 & 3.5 & 0.6 & 63.8 & 24.4 \\
 & IILS & 97.6 & 88.1 & 60.2 & 4.2 & 0.8 & 67.4 & 30.1 \\
\addlinespace[2pt]
CSSE-CSP & Full & 97.5 & 95.9 & 51.9 & 4.0 & 0.7 & 67.6 & 24.9 \\
 & IILS & 98.9 & 95.9 & 77.9 & 1.9 & 1.0 & 61.8 & 42.4 \\
 & IILS (LT) & 98.9 & 96.1 & 77.0 & 1.5 & 0.9 & 73.4 & \textbf{53.0} \\
\midrule
\multicolumn{9}{@{}l}{\textbf{Crystalite (this work) | MP20}} \\
\addlinespace[2pt]
Original representation & Full & 97.5 & 96.4 & 53.3 & 4.7 & 1.0 & 62.0 & 22.2 \\
 & IILS & 98.0 & 96.9 & 56.7 & 3.7 & 1.2 & 69.2 & 32.0 \\
\addlinespace[2pt]
CSSE-CSP & Full & 97.3 & 96.2 & 53.9 & 4.4 & 0.7 & 59.7 & 22.2 \\
 & IILS & 98.2 & 97.6 & 64.7 & 3.5 & 1.2 & 70.3 & 41.8 \\
 & IILS (LT) & 98.2 & 97.6 & 58.7 & 4.1 & 1.3 & 78.3 & \textbf{44.1} \\
\midrule
\multicolumn{9}{@{}l}{\textbf{DiffCSP (this work) | AlexMP20}} \\
\addlinespace[2pt]
One-hot & Full & 95.8 & 95.4 & 60.7 & 1.7 & 0.6 & 67.9 & 37.6 \\
 & IILS & 94.9 & 87.6 & 61.4 & 3.1 & 0.2 & 55.4 & 25.5 \\
\addlinespace[2pt]
CSSE-CSP & Full & 95.6 & 95.5 & 48.7 & 1.6 & 0.5 & 78.5 & 36.2 \\
 & IILS & 98.3 & 96.8 & 84.7 & 0.9 & 0.6 & 81.5 & \textbf{68.9} \\
\midrule
\multicolumn{9}{@{}l}{\textbf{Terminal substitutions (DiffCSP, CSSE-CSP) | MP20}} \\
\addlinespace[2pt]
Codebook substitution & --- & 93.8 & 93.8 & 78.1 & 0.0 & 0.0 & 6.2 & 1.1 \\
Hautier substitution & --- & 99.6 & 99.3 & 84.2 & 0.4 & 0.0 & 18.1 & 9.1 \\
\bottomrule
\end{tabular*}
\begin{tablenotes}[flushleft]
\footnotesize
\item All values are percentages. Reported results are from LeMat-GenBench~\cite{betala2025lemat}, restricted to previously published methods that rank highly on that benchmark. Our results, including ion exchange, are means of three independent sampling runs of 2,500 retained structures, evaluated with LeMat-GenBench after MatterSim-1M relaxation.
Ion exchange starts from known crystals in the MP20 training set.
Metastable includes Stable and is defined by $E_{\mathrm{hull}}\leq0.1\,\mathrm{eV\,atom^{-1}}$. SUN and MSUN additionally require uniqueness and novelty for Stable and Metastable structures, respectively.
DiffCSP uses $\sigma=1.0$ for Full and $\sigma=0.2$ for IILS. Crystalite uses matched sampling settings. LT denotes longer training. Bold marks the highest MSUN within each of our architecture--dataset groups.
\end{tablenotes}
\end{threeparttable}
\end{table}

The frontier expansion is accompanied by higher MSUN on MP20 (Table~\ref{tab:benchmark}).
At the standard training budgets, the CSSE-CSP IILS configurations reach 42.4\% for DiffCSP and 41.8\% for Crystalite, compared with 37.1\% for Creator-MV-DiT~\cite{creator2026creator} and 22.3\% for Chemeleon2~\cite{park2026chemeleon2} in LeMat-GenBench.
Both IILS configurations also exceed the 31.9\% MSUN obtained under our evaluation with template-based ion exchange~\cite{szymanski2025baselines}, which starts from known MP20 training structures.
Longer training yields a larger additional MSUN gain for DiffCSP, which reaches 53.0\%, and a smaller gain for Crystalite, which reaches 44.1\%.

The MSUN advantage over Full persists across the tested architectures and training datasets (Fig.~\ref{fig:performance-transfer}b and Table~\ref{tab:benchmark}).
With CSSE-CSP, MSUN is 17.5 percentage points higher under IILS than Full for DiffCSP on MP20, 19.6 points higher for Crystalite~\cite{veljkovic2026crystalite} on MP20, and 32.7 points higher for DiffCSP on AlexMP20.
The MSUN advantage also extends to the other chemically informed element representations tested here (Table~\ref{tab:benchmark}).
For DiffCSP on MP20, Crystalite-style, CrystalDiT-style and CSSE-MLIP all give higher MSUN under IILS than Full, although the gains are smaller than with CSSE-CSP.
Crystalite likewise achieves higher MSUN under IILS than Full while retaining its original element representation.

The MSUN gains do not require an increase in the overall metastable fraction.
Novelty is higher under IILS in all three CSSE-CSP settings, while both sampling methods retain unique fractions of 95.4--98.1\% (Fig.~\ref{fig:performance-transfer}b).
On MP20, DiffCSP achieves higher MSUN although its metastable fraction falls from 67.6\% to 61.8\%, whereas the Crystalite metastable fraction rises from 59.7\% to 70.3\%.
On AlexMP20, novelty and metastability instead increase together, from 48.7\% to 84.7\% and from 78.5\% to 81.5\%, respectively.

MSUN candidates from both samplers are distributed across multiple regions of the projected MP20 reference distribution (Fig.~\ref{fig:performance-transfer}c).
IILS produces more such candidates in this comparison, with the checkpoint, sampling noise scale and generation budget held fixed (Supplementary Note~\SIref{sec:supp_tsne}).
The MSUN advantage also persists within restricted composition subsets (Fig.~\ref{fig:performance-transfer}d and Supplementary Note~\SIref{sec:supp_composition_counts}).
For DiffCSP with CSSE-CSP on MP20, IILS achieves higher MSUN within each group containing three to six distinct elements, whereas binary crystals give 20.0\% under IILS and 22.2\% under Full.
In the single-run filtering evaluations, the advantage persists after excluding rare-earth elements, restricting structures to at most four distinct elements, or applying both filters.
With both restrictions, MSUN is 33.9\% for IILS and 24.7\% for Full, measured within each retained subset.
The gain is therefore not confined to rare-earth-containing structures or compositions with more than four elements.

\FloatBarrier

\subsection{Crystal discovery under target-property constraints}

\begin{figure}[!tp]
    \centering
    \includegraphics[width=\textwidth]{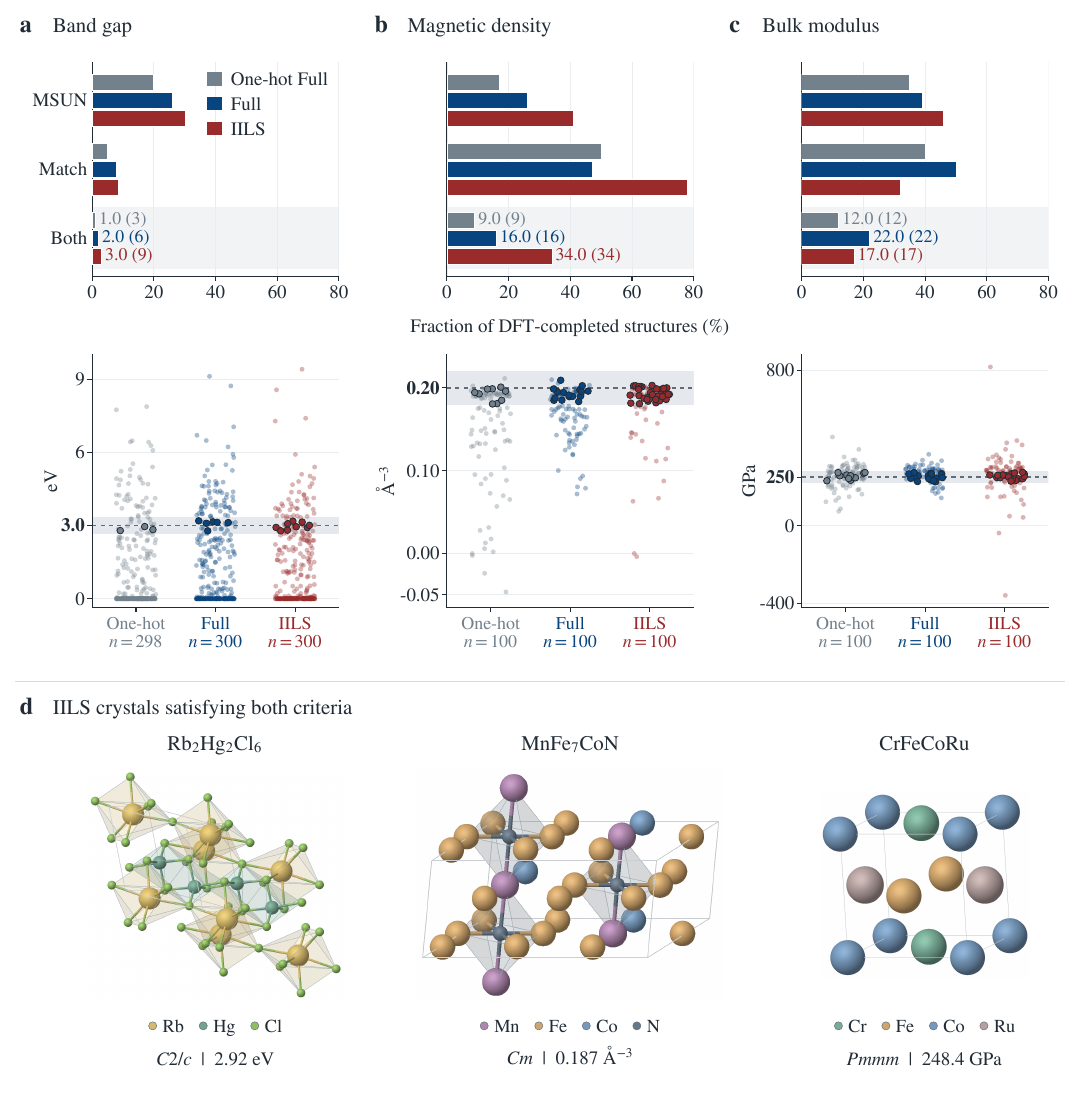}
    \caption[Benefits and limits under property constraints]{%
    \textbf{Benefits and limits of IILS under property constraints.}
DiffCSP models are fine-tuned on AlexMP20.
Full and IILS denote CSSE-CSP unless labelled one-hot.
\textbf{a--c,} Upper plots show MSUN, property matching (Match, within 10\% of the target) and their intersection (Both), as percentages of the evaluated structures.
Parentheses give Both counts.
Lower plots show all plotted DFT property values.
$n$ denotes the number of evaluated structures entering each fraction for each method and task.
Circled points satisfy Both.
Shaded bands and dashed lines mark match windows and targets.
Each configuration includes 300 (\textbf{a}) or 100 (\textbf{b,c}) DFT-evaluated structures.
\textbf{d,} Representative IILS structures satisfying both criteria, with compositions, space groups and DFT properties.
Polyhedra show Rb/Hg--Cl and N-centred coordination environments.
MSUN uses the screening criteria defined in Fig.~\ref{fig:training-discovery}.
DFT evaluates the target properties.
    }
    \label{fig:property-constraints}
\end{figure}

Property-guided discovery requires candidates to satisfy the MSUN and target-property criteria simultaneously.
We test whether retaining intermediate element identities improves this joint outcome for band gap, magnetic density and bulk modulus (Fig.~\ref{fig:property-constraints}a--c).
DiffCSP models with one-hot and CSSE-CSP representations are fine-tuned on property-labelled AlexMP20 structures and sampled using classifier-free guidance~\cite{ho2022cfg} (Supplementary Note~\SIref{sec:supp_conditional_training}).
Full and IILS use the same fine-tuned CSSE-CSP model for each task, without further parameter updates for IILS.
We evaluate the target properties using density functional theory (DFT).
Match denotes the fraction within 10\% of the target, and Both denotes the fraction satisfying this requirement together with the MSUN criteria.
All three fractions are calculated among the evaluated structures of each method.

Under Full sampling, CSSE-CSP gives higher MSUN and Both fractions than one-hot encoding for all three targets (Fig.~\ref{fig:property-constraints}a--c).
IILS further increases all three fractions for band gap and magnetic density.
At the 3.0\,eV band-gap target, Both increases from 2.0\% under Full to 3.0\% under IILS (Fig.~\ref{fig:property-constraints}a).
The improvement is larger for magnetic density, where Both rises from 9.0\% to 34.0\% (Fig.~\ref{fig:property-constraints}b).
The magnetic-density distribution is more concentrated near the target under IILS, whereas the band-gap distributions remain broad under both samplers.

The bulk-modulus task reveals a different balance between discovery and property matching (Fig.~\ref{fig:property-constraints}c).
At a target of 250\,GPa, IILS raises MSUN from 39.0\% to 46.0\%, but Match falls from 50.0\% to 32.0\%.
The fraction satisfying both criteria decreases from 22.0\% to 17.0\%, although it remains above the 12.0\% one-hot baseline.
The benefit of retaining intermediate element identities is therefore task dependent in these comparisons, and higher MSUN alone does not establish an improvement in property-guided discovery.
Despite these task-dependent effects, CSSE-CSP with IILS achieves higher joint yields than the corresponding one-hot baselines across all three property targets (Supplementary Note~\SIref{sec:supp_property_comparison}).

Representative IILS-generated candidates satisfying both the MSUN and target-property criteria are shown in Fig.~\ref{fig:property-constraints}d. 
These include Rb$_2$Hg$_2$Cl$_6$ with a DFT band gap of 2.92\,eV, MnFe$_7$CoN meeting the magnetic-density target, and CrFeCoRu with a DFT bulk modulus of 248.4\,GPa. 

\FloatBarrier

\section{Discussion}

Longer training can improve the stability of generated crystals at the expense of novelty~\cite{veljkovic2026crystalite}.
Our results demonstrate that saturation in the fraction of metastable, unique and novel (MSUN) crystals under full sampling does not preclude further gains from longer training.
With a chemically informed element representation, IILS maintains high novelty and enables additional MSUN gains from later model checkpoints by retaining intermediate element identities.
These findings suggest that intermediate element assignments are not merely transient estimates of terminal identities, but can also support viable crystal configurations that differ from those obtained through complete type denoising.
The value of further training for discovery must therefore be assessed together with the sampling procedure.

These findings motivate the joint design of element representations and sampling procedures for crystal discovery.
The element codebook not only encodes chemical relationships but also shapes which intermediate assignments can be effectively exploited through identity locking. 
Such representations could combine knowledge learned from complementary atomistic tasks and encode how element compatibility depends on local coordination and oxidation state.
Determining which chemical relationships remain informative at different stages of denoising would help tailor these representations to the intermediate element states at which identities are projected.

Adaptive projection could use the evolving element states and local geometry to select when and where to fix identities.
Understanding how training changes the contributions of projected identities and subsequent geometric denoising would help select projection rules that remain effective across checkpoints.
For property-guided generation, incorporating the target property into these decisions could increase the fraction of candidates satisfying both discovery and property criteria.

\section{Methods}

\subsection{Intermediate identity-locking sampling}

We consider crystal diffusion models with continuous element states $\mathbf{a}_{i,t}$, fractional atomic coordinates $\mathbf{f}_{i,t}$ and lattice matrix $\mathbf{l}_t$, where $i$ indexes atomic sites.
Diffusion time decreases from the noise state at $t=1$ to the terminal state at $t=0$.
Given an element codebook $E$ with row vectors $\{\mathbf{e}_z\}_{z\in\mathcal{Z}}$, we define the identity projected from an intermediate element state by
\begin{equation}
\hat z_i(t)=\underset{z\in\mathcal{Z}}{\arg\max}\,
\frac{\mathbf{a}_{i,t}^{\mathsf T}\mathbf{e}_z}
{\|\mathbf{a}_{i,t}\|_2\|\mathbf{e}_z\|_2}.
\label{eq:method_projection}
\end{equation}
This is the nearest-codeword projection under cosine distance.
Full sampling updates all three channels to $t=0$ using the base model's sampler~\cite{jiao2023diffcsp,veljkovic2026crystalite}.
IILS follows the same process until $t_s$, then fixes each element identity to $\hat z_i(t_s)$ and continues denoising only the coordinates and lattice.
The projection itself changes neither coordinates nor lattice, and no model parameters are updated.

For paired trajectory analyses, Full and IILS share the initial state and random history up to $t_s$.
The branch at $t_s$ preserves site correspondence, allowing identities to be compared at the same sites in the paired terminal outputs.

\subsection{Construction of element representations}

Chemical Similarity-based Spherical Encoding (CSSE) constructs a fixed element codebook from representations learned by pretrained atomistic models.
Given atomic representations $\mathbf{h}_i$, we define the prototype of element $z$ and the pairwise cosine-similarity matrix as
\begin{equation}
\bar{\mathbf{h}}_z=\frac{1}{|I_z|}\sum_{i\in I_z}\mathbf{h}_i,
\qquad
R_{zz'}=\frac{\bar{\mathbf{h}}_z^{\mathsf T}\bar{\mathbf{h}}_{z'}}
{\|\bar{\mathbf{h}}_z\|_2\|\bar{\mathbf{h}}_{z'}\|_2},
\label{eq:method_prototypes}
\end{equation}
where $I_z$ denotes the reference atoms of element $z$.
When explicit element embeddings are available, they are used directly as prototypes.
We eigendecompose $R$ and retain its $d$ largest non-negative eigenvalues:
\begin{equation}
R=U\Lambda U^{\mathsf T},
\qquad
X_d=U_d\Lambda_d^{1/2},
\qquad
E=\operatorname{RowNorm}(X_d).
\label{eq:method_csse}
\end{equation}
Spectral truncation approximates the dominant element-similarity structure, while row normalization constrains all codewords to the unit sphere.
We use $d=8$ (Supplementary Note~\SIref{sec:supp_dimension_selection}) and keep the resulting codebook fixed during generative-model training.
Comparisons with alternative codebooks constructed from the same pretrained representations, including direct prototypes and PCA-based encodings, are provided in Supplementary Note~\SIref{sec:supp_codebook_alternatives}.

CSSE-CSP uses element embeddings from a DiffCSP crystal structure prediction model trained on AlexMP20~\cite{jiao2023diffcsp,zeni2025mattergen}.
CSSE-MLIP uses atomic descriptors from the pretrained DPA3 3.1M interatomic potential~\cite{zhang2026dpa3}, averaged by element over the MP20 training set, with pretrained element embeddings used for elements absent from this set.

The comparisons include one-hot encoding, the 16-dimensional PCA representation of Crystalite~\cite{veljkovic2026crystalite}, and a two-dimensional period-and-group representation adapted from CrystalDiT~\cite{yi2026crystaldit}.
Random Spherical uses spherical codewords matched to the CSSE dimensionality and norm without prescribed chemical relationships, whereas Random Permutation reassigns element labels to the CSSE-MLIP codewords while preserving their geometry.
Further construction details are provided in Supplementary Note~\SIref{sec:supp_codebook_sources}.

\subsection{Models, training and sampling settings}

We use the MP20 dataset released with DiffCSP and the AlexMP20 dataset released with MatterGen, retaining their original data splits and preprocessing procedures~\cite{jiao2023diffcsp,zeni2025mattergen}.
Both datasets contain crystals with at most 20 atoms per unit cell.
The base generative models are DiffCSP and Crystalite~\cite{veljkovic2026crystalite}, using their released architectures without modification.
Comparisons between element representations retain the corresponding model architecture and training configuration.
DiffCSP follows its released MP20 training settings, except that the learning-rate scheduler reduction factor is changed from 0.6 to 0.95; the same hyperparameters are used on AlexMP20.
Crystalite follows its released training configuration.

The standard and extended training budgets for DiffCSP are 10,000 and 100,000 epochs on MP20, and 1,000 and 2,500 epochs on AlexMP20, respectively.
For Crystalite on MP20, the corresponding budgets are $2.5\times10^6$ and $2.2\times10^7$ optimisation steps.
Extended training changes only the training duration, without modifying the data, objective or other training settings.
For the unconditional benchmarks, Full sampling uses the checkpoint with the highest MSUN among those evaluated at regular intervals, whereas IILS uses the final checkpoint of the corresponding training run.
The checkpoint evaluation intervals are 1,000 epochs for DiffCSP and $5\times10^5$ optimisation steps for Crystalite.

We denote the element-type Langevin sampling scale by $\sigma$.
For the main DiffCSP benchmarks in Fig.~\ref{fig:performance-transfer}a,b and Table~\ref{tab:benchmark}, as well as the continued-training comparison in Fig.~\ref{fig:training-discovery}, Full uses $\sigma=1$, whereas IILS uses $\sigma=0.2$.
The divergent training trends persist when Full and IILS are compared at a matched sampling scale of $\sigma=0.2$ (Supplementary Note~\SIref{sec:supp_sampling_noise}).
Crystalite uses matched sampling settings for Full and IILS.
For the unconditional benchmarks, DiffCSP uses the representation-specific projection times indicated in Fig.~\ref{fig:representation-timing}b, whereas Crystalite uses $t_s=0.2$ for both element representations.
In the DiffCSP projection-time scans (Fig.~\ref{fig:representation-timing}e), $t_s$ ranges from 0 to 0.3 in increments of 0.05, with $\sigma$ fixed across projection times at 1 for one-hot and 0.2 for CSSE-CSP, including the corresponding Full references.
All other sampling parameters follow the respective official implementations.

\subsection{Element-replacement controls}

The Projected identities + Full geometry control combines element identities $\hat z_i(t_s)$ projected from a Full trajectory at $t_s=0.2$ with the terminal coordinates and lattice of the same trajectory at $t=0$.
The resulting structure is relaxed directly, without additional diffusion.

Static substitution controls start from Full terminal structures and randomly replace a fraction of atomic sites approximately matched to the terminal type-change fraction under IILS (Supplementary Note~\SIref{sec:supp_substitution_methods}).
The Codebook control samples replacement elements according to cosine similarities between CSSE-CSP codewords, whereas the Hautier control uses empirical ionic-substitution probabilities~\cite{hautier2011substitution}.

To examine the effect of geometric adaptation, the coordinates and lattice of substituted structures are re-noised to $t=0.2$ using the base model's forward diffusion processes, then denoised to $t=0$ with element identities fixed while coordinates and lattice evolve, followed by relaxation.

Template-based ion exchange is evaluated separately as a discovery benchmark~\cite{szymanski2025baselines}.
We use the \texttt{matgen\_baselines} implementation with parent structures drawn from the 27,136-crystal MP20 training split, restricted to those with reported energy above hull at most $0.1\,\mathrm{eV\,atom^{-1}}$.
Generated structures undergo the same relaxation and evaluation as the diffusion-model benchmarks (Supplementary Note~\SIref{sec:supp_template_baselines}).

\subsection{Candidate compatibility and paired-output analysis}

We quantify empirical element-substitution associations using the Hautier statistics~\cite{hautier2011substitution}.
For ionic species $a^{q_a}$ and $b^{q_b}$, the association score and its element-level counterpart are
\begin{align}
\rho(a^{q_a},b^{q_b})
&=\frac{P(a^{q_a},b^{q_b})}{P(a^{q_a})P(b^{q_b})},\\
\rho(a,b)
&=\max_{q_a\in\mathcal O_a,\,q_b\in\mathcal O_b}
\rho(a^{q_a},b^{q_b}),
\label{eq:method_association}
\end{align}
where $\mathcal O_z$ is the set of commonly observed oxidation states of element $z$.
The element-level score is set to zero if either set is empty.

For a codebook containing $K$ elements, let $r_a(b)$ be the ascending rank of $\rho(a,b)$ among the other $K-1$ elements, numbered from 0 to $K-2$.
We symmetrize the normalized ranks as
\begin{equation}
H_{ab}=\frac{1}{2}\left[\frac{r_a(b)}{K-2}+\frac{r_b(a)}{K-2}\right].
\end{equation}
Let $z_{i1},z_{i2},z_{i3}$ denote the three nearest codewords to the intermediate element state at site $i$.
The candidate-association score is
\begin{equation}
C=\frac{1}{3N}\sum_{i=1}^{N}\sum_{1\leq j<k\leq3}H_{z_{ij}z_{ik}},
\label{eq:method_candidate_rank}
\end{equation}
where $N$ is the number of analysed sites.
Higher $C$ indicates stronger empirical substitution associations among candidate elements.
Trajectory selection, candidate-frequency analysis and statistical reporting are detailed in Supplementary Note~\SIref{sec:supp_candidate_analysis}.

For paired terminal outputs, let $d_i=\mathbb{I}[z_i^{\mathrm{IILS}}\neq z_i^{\mathrm{Full}}]$.
A differing element pair is considered compatible when the two elements share at least one oxidation state and their association score satisfies $\rho\geq\tau=3.40$.
The threshold is the 85th percentile of $\rho$ over ordered codebook element pairs $(a,b)$ with $a\neq b$, including pairs with zero association scores.
Denoting the corresponding compatibility indicator by $k_i$, we define
\begin{equation}
A_{\mathrm{IILS,Full}}=\frac{1}{N}\sum_i d_i,
\qquad
Q_{\mathrm{IILS,Full}}=\frac{1}{N}\sum_i d_i k_i.
\label{eq:method_aq}
\end{equation}
Both quantities are normalized by the total number of atomic sites across paired samples.
Thus, $(A-Q)_{\mathrm{IILS,Full}}$ measures identity changes that do not satisfy the empirical compatibility criterion, while $(Q/A)_{\mathrm{IILS,Full}}$, for $A>0$, gives the compatible fraction among changed sites.
This criterion describes empirical element-pair compatibility rather than stability at a particular crystal site.

At the crystal level, let $m_b^{\mathrm{IILS}}$ and $m_b^{\mathrm{Full}}$ indicate whether the corresponding outputs in pair $b$ satisfy the MSUN criteria.
For $B$ paired outputs, we define gain and loss as
\begin{equation}
G=\frac{1}{B}\sum_b(1-m_b^{\mathrm{Full}})m_b^{\mathrm{IILS}},
\qquad
L=\frac{1}{B}\sum_b m_b^{\mathrm{Full}}(1-m_b^{\mathrm{IILS}}).
\label{eq:method_gain_loss}
\end{equation}
The net MSUN change is $100(G-L)$ in percentage points.
Additional compatibility and gain--loss analyses are reported in Supplementary Note~\SIref{sec:supp_compatibility_gains}.

\subsection{Crystal evaluation and statistical reporting}

For the repeated unconditional benchmarks, each run contains 2,500 retained structures after excluding compositions containing noble gases or radioactive elements.
Generated structures are relaxed with MatterSim-1M~\cite{yang2024mattersim} and evaluated using LeMat-GenBench~\cite{betala2025lemat}.
The benchmark assesses structural and compositional validity, uniqueness, novelty relative to the LeMat-Bulk reference database, energetic stability, and additional distributional and diversity metrics (Supplementary Note~\SIref{sec:supp_evaluation_metrics}).
Energy-above-hull estimates are obtained through the benchmark's multi-MLIP evaluation procedure against reference convex hulls, using MLIPs distinct from MatterSim-1M, which is employed for structural relaxation.
We define MSUN as the joint yield of structures satisfying metastability ($E_{\mathrm{hull}}\leq0.1\,\mathrm{eV\,atom^{-1}}$), uniqueness and novelty, combining the stable and metastable-only categories reported separately by LeMat-GenBench.

External benchmark results are taken directly from LeMat-GenBench without re-evaluation.
For the projection-time scans, reported fractions use the evaluated sample count recorded for each run.
Additional results obtained using the MatterGen evaluation pipeline are reported in Supplementary Note~\SIref{sec:supp_alternative_evaluation}.
All training, generation and evaluation, except DFT calculations, are performed on a single NVIDIA A800 GPU.
Software dependencies and version information are provided in the accompanying code repository.

\subsection{Property-guided generation and DFT evaluation}

Property-guided DiffCSP models are initialized from the unconditional one-hot and CSSE-CSP models trained on AlexMP20.
Each scalar property is standardized using the training-set mean and standard deviation, encoded by a sinusoidal embedding followed by a two-layer multilayer perceptron, and added to the diffusion-time embedding.
The property branch is initialized with zero output.
During fine-tuning, property conditions are independently dropped with probability 0.2, and structures without property labels are treated as unconditional.
The original denoising objective is retained, and the checkpoint with the lowest validation loss is selected.
Further architectural and optimization details are provided in Supplementary Note~\SIref{sec:supp_conditional_training}.

Property-guided sampling uses classifier-free guidance (CFG)~\cite{ho2022cfg}, with a guidance scale of $w=2$ applied to the lattice, fractional-coordinate and element-type channels (Supplementary Note~\SIref{sec:supp_conditional_training}).
Full and IILS use the same fine-tuned CSSE-CSP checkpoint for each task.
The AlexMP20 band-gap and magnetic-density labels are obtained from DFT calculations, whereas its bulk-modulus labels come from a machine-learning predictor.
All three properties of generated structures are evaluated using DFT.

Following MatterSim-1M relaxation, DFT calculations are performed using VASP 6.3.0~\cite{kresse1996vasp}, the projector-augmented-wave method~\cite{blochl1994paw}, and the PBE exchange--correlation functional~\cite{perdew1996pbe}.
Materials Project parameter sets are used, including spin-polarization and DFT+$U$ settings.
Band gaps are obtained from non-self-consistent band-structure calculations, magnetic density from total magnetization divided by cell volume, and bulk moduli from the Voigt--Reuss--Hill averages of calculated elastic tensors.
Detailed computational workflows are provided in Supplementary Note~\SIref{sec:supp_dft_validation}.

For each configuration, up to 300 structures for band gap and 100 structures each for magnetic density and bulk modulus completed DFT evaluation.
A generated structure is considered property-matched if its evaluated property lies within $\pm 10\%$ of the target value.
Match denotes the fraction of property-matched structures, and Both denotes the fraction satisfying both the property-matching and MSUN criteria.
All reported fractions are calculated over these DFT-evaluated structures.
Supplementary Note~\SIref{sec:supp_property_comparison} provides the target values, completed sample counts and comparisons with pretrained MatterGen property-conditional models.

\section{Data availability}

The repository contains all sampled crystal structures used in the benchmarking analyses presented in this work.
The MP-20 dataset used for training is available at \url{https://github.com/txie-93/cdvae}, and the Alex-MP-20 dataset was obtained from the MatterGen repository at \url{https://github.com/microsoft/mattergen}.

\section{Code availability}

The source code is publicly available at \url{https://github.com/beform88/CSSE-IILS-Crystal} and includes the implementations of IILS within the DiffCSP and Crystalite frameworks, the CSSE codebook and the property-guided module.

\section{Acknowledgements}

We acknowledge the AI for Science Institute for providing the computational resources used in this work.
H.W. acknowledges support from the National Natural Science Foundation of China (Grants No.~12525113 and No.~12561160120).

\ifdefined\arxivbuild
\documentclass[12pt]{article}

\usepackage[T1]{fontenc}
\usepackage[utf8]{inputenc}
\usepackage{lmodern}
\usepackage[a4paper,margin=1in]{geometry}
\usepackage{amsmath,amssymb,bm}
\usepackage{graphicx}
\graphicspath{{figures/}}
\usepackage{booktabs,multirow,array,makecell,threeparttable}
\usepackage[numbers,sort&compress]{natbib}
\usepackage{microtype}
\usepackage[hidelinks]{hyperref}
\usepackage{float,placeins}

\title{Supplementary Information\\[0.5em]
\large Retaining intermediate element identities enables longer training to improve crystal discovery}
\author{}
\date{}

\begin{document}
\ifdefined\arxivbuild
  \clearpage
  \nolinenumbers
  \graphicspath{{}}
  \setcounter{secnumdepth}{3}
  \setcounter{section}{0}
  \setcounter{subsection}{0}
  \setcounter{figure}{0}
  \setcounter{table}{0}
  \setcounter{equation}{0}
  \renewcommand{\theHsection}{suppsec.\arabic{section}}
\fi

\renewcommand{\thesection}{S\arabic{section}}
\renewcommand{\thefigure}{S\arabic{figure}}
\renewcommand{\thetable}{S\arabic{table}}
\renewcommand{\theequation}{S\arabic{equation}}
\renewcommand{\theHfigure}{suppfig.\arabic{figure}}
\renewcommand{\theHtable}{supptab.\arabic{table}}
\renewcommand{\theHequation}{suppeq.\arabic{equation}}

\ifdefined\arxivbuild
\tableofcontents
\clearpage
\begin{center}
{\LARGE Supplementary Information\par}
\end{center}
\else
\maketitle
\tableofcontents
\clearpage
\fi

\section{Element representations and CSSE construction}
\label{sec:supp_codebooks}

\subsection{Codebook sources and construction}
\label{sec:supp_codebook_sources}

This section describes the sources and construction of the element codebooks used in this study.

For manually constructed chemistry-aligned codebooks, Crystalite-style encoding uses the 16-dimensional PCA element representation adopted in the original Crystalite model.
CrystalDiT-style encoding uses the two-dimensional element representation constructed from periodic-table period and group information in CrystalDiT.
Both representations are taken from their respective official implementations.

To distinguish the effects of continuous codebook geometry from those of specific element relationships, we construct two control representations: Random Spherical and Random Permutation.
Random Spherical uses randomly generated spherical codewords with the same dimensionality and scale as CSSE, without incorporating learned or manually specified chemical relationships between elements.
Random Permutation randomly reassigns element identities to the codewords of CSSE-MLIP while preserving the codeword set and its overall geometry.
The random-control results reported in the main text are averaged over three independent random seeds.

CSSE-MLIP uses the pretrained DPA3 3.1M model as its source model, with the MP20 training set as the reference structure set.
Atomic-level representations are extracted using \texttt{DeepPot.eval\_descriptor} provided by DeePMD-kit~\cite{zeng2023deepmd} and aggregated by element type to obtain element prototypes.
For elements absent from the MP20 reference structures, the corresponding pretrained type embeddings are used as prototypes.
The resulting prototypes are used to construct the element codebook following the CSSE procedure described in Methods.

CSSE-CSP uses pretrained type embeddings from a DiffCSP crystal structure prediction model trained on AlexMP20 with the official MP20 training configuration~\cite{jiao2023diffcsp}.
The pretrained type embeddings are directly used as element prototypes.
The subsequent codebook construction follows the same CSSE procedure as CSSE-MLIP.

All codebooks remain fixed throughout subsequent crystal generative model training.

\subsection{Comparison with direct prototypes and PCA}
\label{sec:supp_codebook_alternatives}
To evaluate the contribution of spectral reconstruction, we build three codebooks from the same source representation(pretrained DPA3 3.1M): CSSE-MLIP, the norm-normalised element prototypes used directly as codewords, and an eight-dimensional PCA codebook.
PCA preserves the principal variance directions of pretrained representations, but the resulting codebook geometry is also influenced by non-angular information.
Supplementary Table~\ref{tab:supp_csse_alternatives} reports their full-sampling and IILS performance.

CSSE-MLIP achieves the highest MSUN under both Full and IILS among codebooks constructed from the same pretrained representations (Supplementary Table~\ref{tab:supp_csse_alternatives}). 
Under Full sampling, CSSE-MLIP reaches 24.4\% MSUN, compared with 21.5\% for PCA and 22.7\% for directly normalized prototypes. 
Under IILS, the advantage becomes more pronounced: CSSE-MLIP reaches 30.1\%, compared with 25.4\% and 20.4\%, respectively. 
These results show that CSSE provides a more effective element representation for crystal generation under both sampling procedures, with an additional advantage when intermediate identities are fixed.

\begin{table}[htbp]
    \centering
    \caption{\textbf{CSSE, direct pretrained prototypes and PCA under full sampling and IILS.}}
    \label{tab:supp_csse_alternatives}
    \small
    \resizebox{\textwidth}{!}{%
    \begin{tabular}{llrrrrrrr}
        \toprule
        Representation &
        Sampling &
        Novel (\%) &
        Metastable (\%) &
        Stable (\%) &
        MSUN (\%) &
        SUN (\%) &
        $E_{\mathrm{hull}}$ (eV/atom) &
        RMSD (\AA) \\
        \midrule
        CSSE ($d=8$)                 & Full             & 54.92 & 63.80 &  3.48 & 24.40 & 0.64 & 0.09 & 0.21 \\
        CSSE ($d=8$)                 & IILS ($t_s=0.15$) & 60.20 & 67.40 &  4.16 & 30.12 & 0.76 & 0.08 & 0.24 \\
        PCA ($d=8$)                  & Full             & 47.64 & 68.88 &  5.28 & 21.52 & 0.88 & 0.08 & 0.18 \\
        PCA ($d=8$)                  & IILS ($t_s=0.15$) & 57.32 & 65.12 &  4.80 & 25.40 & 0.72 & 0.09 & 0.25 \\
        Pretrained prototype (norm.) & Full             & 61.36 & 53.56 &  3.72 & 22.68 & 0.84 & 0.11 & 0.21 \\
        Pretrained prototype (norm.) & IILS ($t_s=0.15$) & 78.00 & 32.80 &  1.36 & 20.36 & 0.44 & 0.16 & 0.69 \\
        \bottomrule
    \end{tabular}%
    }
\end{table}

\subsection{CSSE dimension selection}
\label{sec:supp_dimension_selection}

We compare the generation performance across different dimensions of CSSE, taking CSSE-MLIP, the codebook built from the frozen DPA3 3.1M encoder, as the representative instance.
Supplementary Table~\ref{tab:supp_csse_dimension} reports the corresponding metrics under full sampling and IILS.

\begin{table}[htbp]
    \centering
    \caption{\textbf{Performance across CSSE dimensions under full sampling and IILS.}}
    \label{tab:supp_csse_dimension}
    \small
    \resizebox{\textwidth}{!}{%
    \begin{tabular}{llrrrrrrr}
        \toprule
        Representation &
        Sampling &
        Novel (\%) &
        Metastable (\%) &
        Stable (\%) &
        MSUN (\%) &
        SUN (\%) &
        $E_{\mathrm{hull}}$ (eV/atom) &
        RMSD (\AA) \\
        \midrule
        CSSE ($d=2$)  & Full              & 62.80 & 47.16 & 2.40 & 19.44 & 0.20 & 0.15 & 0.31 \\
        CSSE ($d=2$)  & IILS ($t_s=0.15$) & 83.59 & 16.10 & 0.15 &  5.52 & 0.00 & 0.32 & 0.76 \\
        CSSE ($d=4$)  & Full              & 55.24 & 63.56 & 4.36 & 23.52 & 0.48 & 0.09 & 0.19 \\
        CSSE ($d=4$)  & IILS ($t_s=0.15$) & 69.80 & 52.24 & 2.44 & 25.08 & 0.40 & 0.12 & 0.33 \\
        CSSE ($d=8$)  & Full              & 54.92 & 63.80 & 3.48 & 24.40 & 0.64 & 0.09 & 0.21 \\
        CSSE ($d=8$)  & IILS ($t_s=0.15$) & 60.20 & 67.40 & 4.16 & 30.12 & 0.76 & 0.08 & 0.24 \\
        CSSE ($d=16$) & Full              & 48.20 & 69.52 & 5.04 & 22.04 & 0.84 & 0.08 & 0.17 \\
        CSSE ($d=16$) & IILS ($t_s=0.15$) & 56.96 & 66.44 & 5.64 & 26.08 & 0.64 & 0.08 & 0.23 \\
        \bottomrule
    \end{tabular}%
    }
\end{table}

The highest MSUN yield is obtained at $d=8$ under both sampling settings, and this dimension is used as the default CSSE dimensionality.

\section{Intermediate-state analysis and IILS sampling}
\label{sec:supp_intermediate}

\subsection{Intermediate-state candidate analysis}
\label{sec:supp_candidate_analysis}

For each model, 1,024 full-sampling crystal generation trajectories are used for intermediate-state analysis.
At each analysed diffusion time, candidate elements are ranked by cosine distance between the continuous element-type state and the fixed codebook.
The three nearest codewords define the candidate neighbourhood, whereas the nearest codeword alone determines the identity used by IILS.

For the chlorine analysis in Fig.~3a, we select sites whose terminal element under Full sampling is Cl.
Candidate frequencies are calculated over the three nearest codewords at each selected site and normalized by the total number of candidate slots, $3N_{\mathrm{Cl}}$.

For the candidate-association analysis in Fig.~3b, the association score $C$ is calculated as defined in Methods.
The reported intervals are the mean plus or minus $1.96s/\sqrt{N}$, where $s$ is the sample standard deviation of site-level association scores and $N$ is the number of analysed sites.
The random reference is estimated from $2\times10^5$ triplets sampled uniformly with replacement from the element codebook.
Its mean is $0.495$.

\subsection{Projection-time sensitivity and operating points}
\label{sec:supp_projection_time}

\begin{figure}[htbp]
    \centering
    \includegraphics[width=\textwidth]{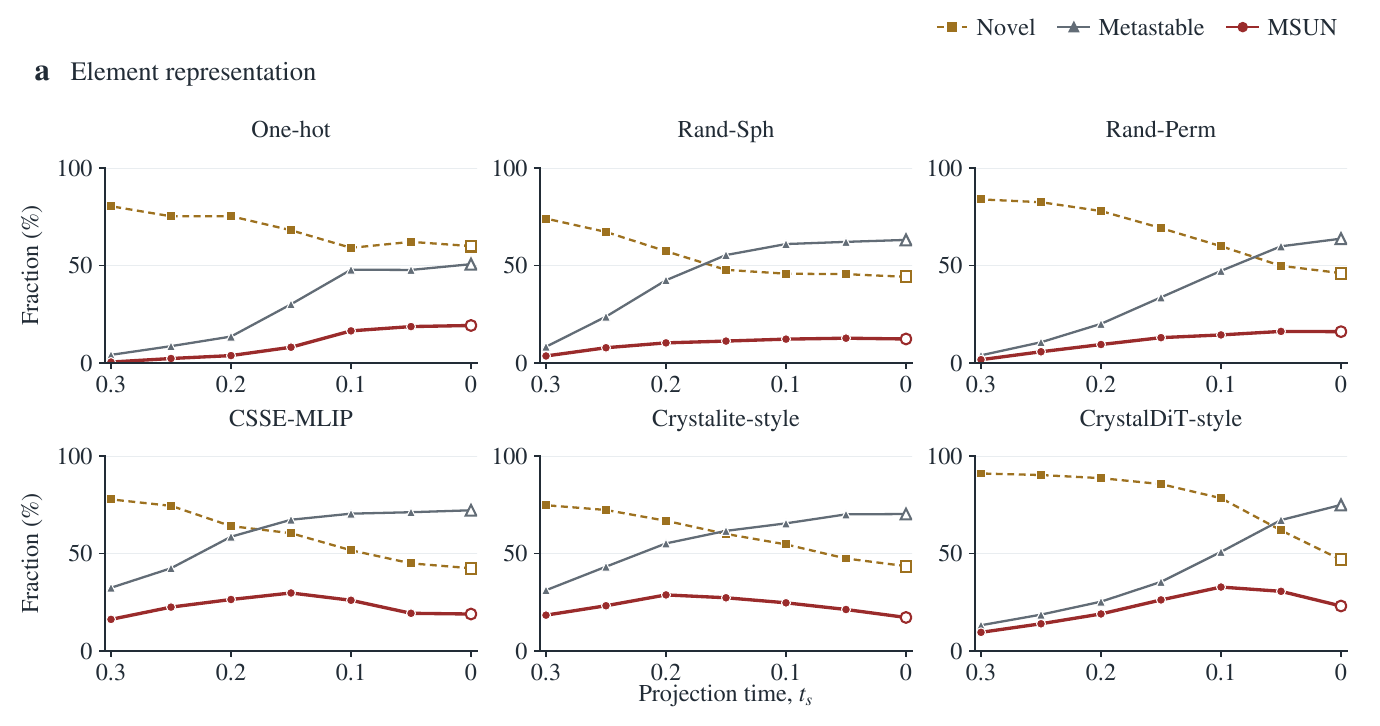}
    \caption{\textbf{Projection-time sensitivity across element representations.}
    Novel, Metastable and MSUN yields versus IILS projection time for DiffCSP with one-hot, random spherical, random permutation, CSSE-MLIP, Crystalite-style and CrystalDiT-style element representations.}
    \label{fig:supp_projection_time_representation}
\end{figure}

\begin{figure}[htbp]
    \centering
    \includegraphics[width=\textwidth]{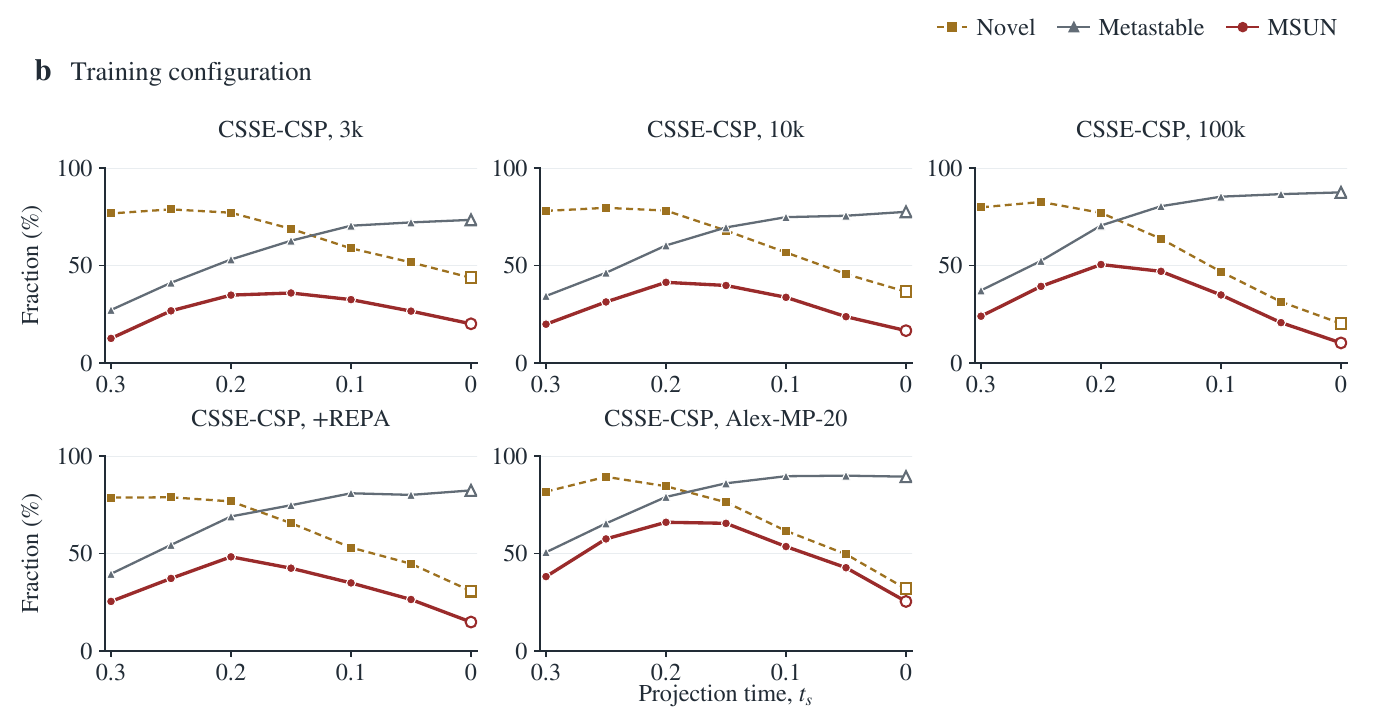}
    \caption{\textbf{Projection-time sensitivity across training configurations.}
    CSSE-CSP with training lengths of 3,000, 10,000 and 100,000 epochs, on the Alex-MP20 dataset, and with the auxiliary REPA objective.}
    \label{fig:supp_projection_time_training}
\end{figure}

\begin{figure}[htbp]
    \centering
    \includegraphics[width=0.685\textwidth]{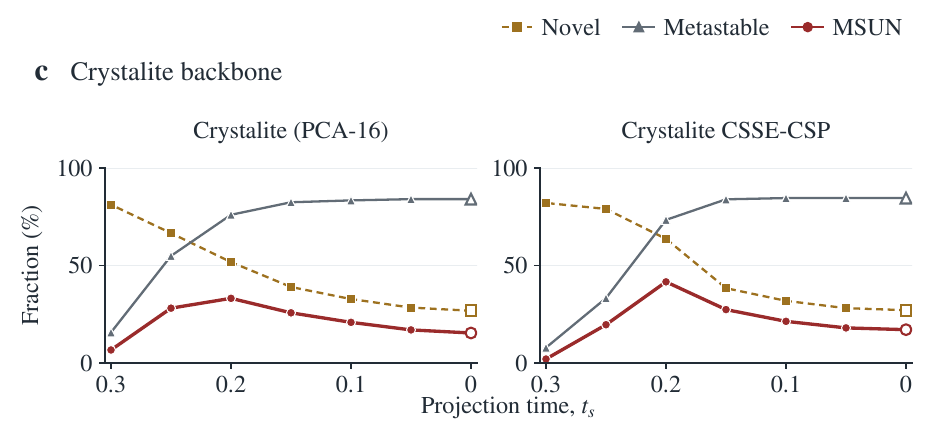}
    \caption{\textbf{Projection-time sensitivity of the Crystalite backbone.}
    Crystalite with PCA-16, the original representation of Crystalite in the main text, and with the CSSE-CSP element representation.}
    \label{fig:supp_projection_time_crystalite}
\end{figure}

Supplementary Figs.~\ref{fig:supp_projection_time_representation}, \ref{fig:supp_projection_time_training} and \ref{fig:supp_projection_time_crystalite} scan the IILS projection time while varying the element representation, the training configuration and the backbone, respectively.
Each experiment is evaluated at $t_{\mathrm{s}}\in\{0,0.05,0.10,0.15,0.20,0.25,0.30\}$, and all panels share the same axes.
The point $t_{\mathrm{s}}=0$ corresponds to sampling without projection and is drawn with hollow markers.
For chemistry-aligned codebooks, MSUN yield varies non-monotonically with projection time and generally peaks at intermediate values.

\subsection{Partial projection and fixed-site fraction}
\label{sec:supp_partial_projection}

\begin{figure}[!tp]
    \centering
    \includegraphics[width=\textwidth]{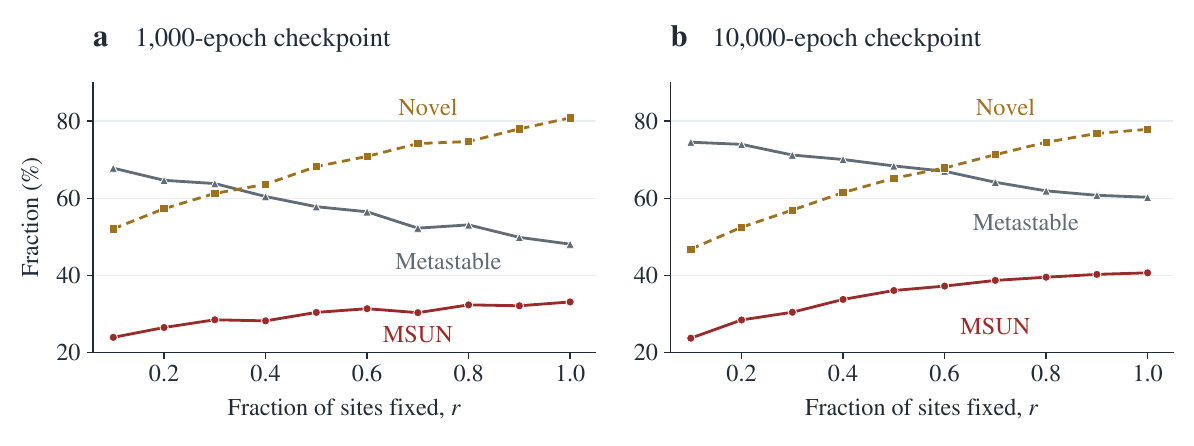}
    \caption{\textbf{Partial identity locking at the 1,000- and 10,000-epoch checkpoints.}
    Metastable, Novel and MSUN versus the fraction $r$ of randomly selected sites whose identities are fixed at $t_s=0.2$, using DiffCSP with CSSE-CSP on MP20 ($\sigma=0.2$).
    \textbf{a,} 1,000-epoch checkpoint.
    \textbf{b,} 10,000-epoch checkpoint.
    Each point uses one sampling run, with fractions calculated over that run's evaluated population ($n=1{,}766$--$2{,}500$).
    Metastable and MSUN include stable structures.}
    \label{fig:supp_partial_projection}
\end{figure}

In the main experiments, IILS projects and fixes the element type of every atomic site at $t_s$, corresponding to a fixed-site fraction $r=1$.
In partial projection, only a fraction $r$ of the sites is projected at $t_s$ and held fixed, while the element types of the remaining sites continue to be updated jointly with the fractional coordinates and the lattice.
The fixed sites are selected uniformly at random per structure.
Supplementary Fig.~\ref{fig:supp_partial_projection} reports the Novel, Metastable and MSUN yields as functions of $r$ at the 1,000- and 10,000-epoch checkpoints, using DiffCSP trained on MP20 with the CSSE-CSP codebook at the main-text operating point ($t_s=0.2$, $\sigma=0.2$), with all other sampling, relaxation and evaluation settings held fixed.
The two checkpoints give consistent trends over $r$.

\subsection{Type-channel Langevin step-size sensitivity}
\label{sec:supp_sampling_noise}

\begin{figure}[htbp]
    \centering
    \includegraphics[width=0.75\textwidth]{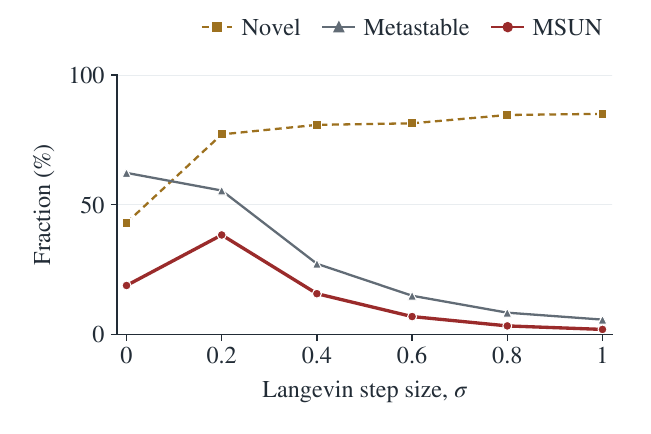}
    \caption{\textbf{Choice of the type-channel Langevin step size in IILS.}
    Metastable, MSUN and Novel fractions of IILS-generated structures as a function of the type-channel Langevin step size $\sigma$ (DiffCSP, MP20, CSSE-CSP).}
    \label{fig:supp_langevin_step}
\end{figure}

IILS uses $\sigma=0.2$, which achieves the highest MSUN yield among the tested type-channel Langevin step sizes (Supplementary Fig.~\ref{fig:supp_langevin_step}). Crystalite retains the default settings of its official implementation for both Full and IILS.

The comparison in Fig.~1 of the main text changes both the type-channel step size $\sigma$ and the sampling procedure. 
To isolate the effect of identity locking, we conduct a longitudinal comparison across 100 paired checkpoints from 1,000 to 100,000 epochs, using the same $\sigma=0.2$ for Full and IILS.
All other sampling settings are matched.

\begin{figure}[htbp]
    \centering
    \includegraphics[width=\textwidth]{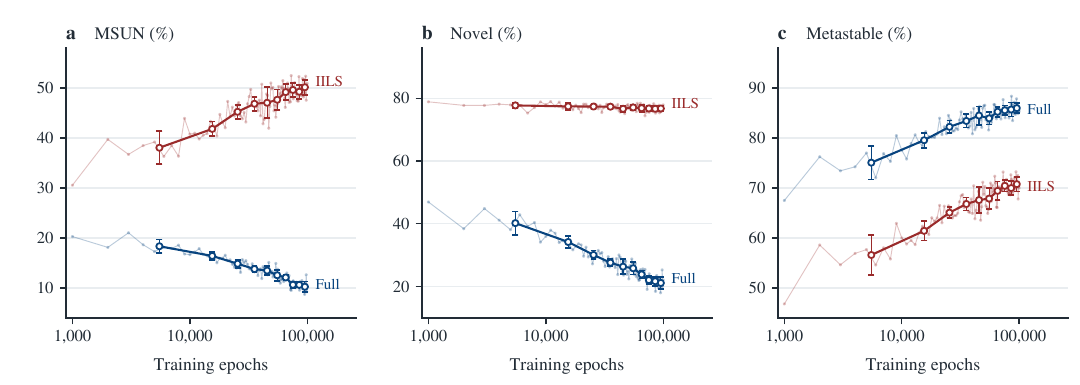}
    \caption{\textbf{Longitudinal control at a matched type-channel step size (DiffCSP, MP20, CSSE-CSP).}
    Both settings use the same 100 training checkpoints, element codebook, random seed and type-channel step size $\sigma=0.2$, and differ only in whether the element identities are projected and fixed at $t_s=0.2$.
    \textbf{a,} MSUN.
    \textbf{b,} Novel.
    \textbf{c,} Metastable.
    Small markers show individual checkpoints at intervals of 1,000 epochs; large open markers with error bars are the mean and standard deviation over consecutive groups of ten checkpoints.
    Blue, full sampling; red, IILS.
    MSUN and Metastable follow the composite definitions of Table~1.}
    \label{fig:supp_matched_sigma}
\end{figure}

Under this matched setting, Full MSUN decreases from 20.3\% to 10.4\% with continued training, whereas IILS MSUN increases from 30.6\% to 50.9\% (Supplementary Fig.~\ref{fig:supp_matched_sigma}a). 
The divergence is primarily associated with novelty: under Full sampling, the Metastable fraction increases from 67.5\% to 85.3\% while Novel decreases from 46.9\% to 22.1\%; under IILS, Metastable increases from 46.8\% to 70.0\% while Novel remains between 76\% and 79\% (Supplementary Fig.~\ref{fig:supp_matched_sigma}b,c).
These results show that the contrasting MSUN trends under continued training persist when the type-channel step size is held fixed.

\subsection{Compatibility and gain--loss analyses}
\label{sec:supp_compatibility_gains}

Supplementary Tables~\ref{tab:supp_projection_csse} and~\ref{tab:supp_projection_onehot} give the Metastable and MSUN gain--loss decomposition over the projection-time scan for CSSE-CSP and one-hot encoding, with $A$, $Q$ and $Q/A$ defined as in Methods.
Because $A$ and $Q$ share the all-site denominator, the tabulated entries satisfy $Q=A\,(Q/A)$.
All gain, loss and net-change values are in percentage points, and the same conventions hold in Supplementary Table~\ref{tab:supp_representation_gainloss}.

\begin{table}[htbp]
    \centering
    \caption{\textbf{Projection-time dependence of gain--loss statistics for CSSE-CSP.}}
    \label{tab:supp_projection_csse}
    \small
    \resizebox{\textwidth}{!}{%
    \begin{tabular}{rrrrrrrrrr}
        \toprule
        $t_{\mathrm{s}}$ &
        Metastable gain &
        Metastable loss &
        MSUN gain &
        MSUN loss &
        $\Delta$Metastable &
        $\Delta$MSUN &
        $A$ &
        $Q$ &
        $Q/A$ \\
        \midrule
        0.05 & 13.92 & 15.52 & 18.28 & 11.84 &  -1.60 &  6.44 &  7.63 &  5.69 & 74.51 \\
        0.10 & 15.68 & 14.00 & 26.28 & 10.48 &   1.68 & 15.80 & 19.09 & 11.90 & 62.34 \\
        0.15 & 12.52 & 20.16 & 31.92 & 10.60 &  -7.64 & 21.32 & 34.81 & 18.40 & 52.86 \\
        0.20 &  9.87 & 25.99 & 35.00 & 10.33 & -16.12 & 24.67 & 48.81 & 20.21 & 41.41 \\
        0.25 &  6.38 & 38.78 & 26.81 & 11.19 & -32.40 & 15.62 & 61.26 & 19.51 & 31.84 \\
        0.30 &  3.88 & 51.29 & 17.88 & 11.41 & -47.41 &  6.47 & 71.75 & 17.92 & 24.97 \\
        \bottomrule
    \end{tabular}%
    }
\end{table}

\begin{table}[htbp]
    \centering
    \caption{\textbf{Projection-time dependence of gain--loss statistics for one-hot encoding.}}
    \label{tab:supp_projection_onehot}
    \small
    \resizebox{\textwidth}{!}{%
    \begin{tabular}{rrrrrrrrrr}
        \toprule
        $t_{\mathrm{s}}$ &
        Metastable gain &
        Metastable loss &
        MSUN gain &
        MSUN loss &
        $\Delta$Metastable &
        $\Delta$MSUN &
        $A$ &
        $Q$ &
        $Q/A$ \\
        \midrule
        0.05 & 12.76 & 12.96 & 10.56 & 10.72 &  -0.20 &  -0.16 &  1.24 &  0.27 & 21.48 \\
        0.10 & 11.96 & 13.92 & 10.16 & 10.64 &  -1.96 &  -0.48 &  4.19 &  0.75 & 17.80 \\
        0.15 &  4.88 & 28.48 &  6.08 & 13.00 & -23.60 &  -6.92 & 20.02 &  2.55 & 12.73 \\
        0.20 &  1.17 & 48.29 &  2.70 & 14.37 & -47.12 & -11.67 & 47.21 &  5.20 & 11.01 \\
        0.25 &  1.16 & 56.11 &  1.17 & 11.73 & -54.95 & -10.56 & 67.26 &  7.13 & 10.60 \\
        0.30 &  0.65 & 59.43 &  0.78 & 11.76 & -58.78 & -10.98 & 78.94 &  8.59 & 10.88 \\
        \bottomrule
    \end{tabular}%
    }
\end{table}

\begin{figure}[htbp]
    \centering
    \includegraphics[width=\textwidth]{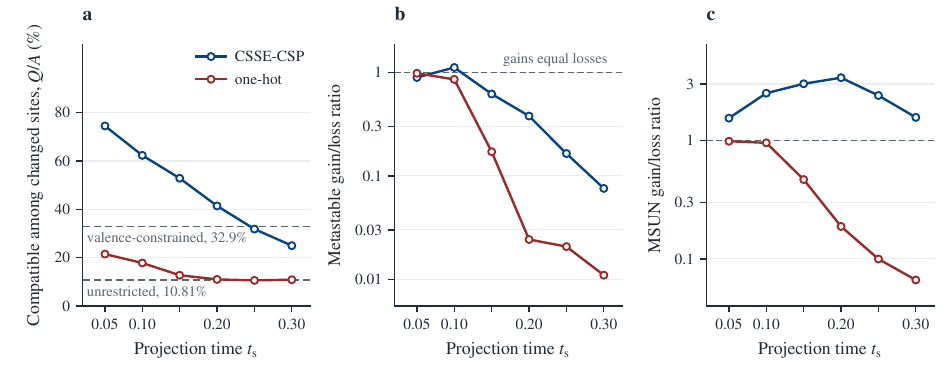}
    \caption{\textbf{Compatibility and gain/loss balance of projected element identities (DiffCSP, MP20).}
    \textbf{a,} Compatible fraction among changed sites, $Q/A$, against the projection time $t_{\mathrm{s}}$; dashed lines mark the compatibility of the two random-substitution baselines, unrestricted ($10.81\%$) and valence-constrained ($32.9\%$).
    \textbf{b,c,} Ratio of metastable (\textbf{b}) and MSUN (\textbf{c}) gain to loss, the number of qualifying outputs created per one destroyed, against $t_{\mathrm{s}}$; dashed lines mark a ratio of one and the vertical axes are logarithmic.
    All values are derived from Supplementary Tables~\ref{tab:supp_projection_csse} and~\ref{tab:supp_projection_onehot}.}
    \label{fig:supp_qa_baseline}
\end{figure}

To contextualize $Q/A$, we consider two random-substitution references: unrestricted element-pair sampling, with 10.81\% compatible pairs, and valence-constrained sampling, with 32.9\%. 
CSSE-CSP exceeds the valence-constrained reference up to $t_{\mathrm{s}}=0.20$, whereas one-hot remains below it throughout the scan (Supplementary Fig.~\ref{fig:supp_qa_baseline}a).

Supplementary Fig.~\ref{fig:supp_qa_baseline}b,c shows the ratio of gains to losses for metastability and MSUN across projection times. 
For metastability, losses dominate at later projection times under both representations, but are substantially larger relative to gains under one-hot encoding. 
For MSUN, CSSE-CSP maintains a gain-to-loss ratio above one throughout the scan, whereas one-hot remains below one.

Supplementary Table~\ref{tab:supp_representation_gainloss} reports the gain--loss decomposition across element representations.
These entries are the separately reported operating-point summaries also used for the compatibility panels of the main text (Fig.~3c,d), so the projection time and the type-channel noise scale $\sigma$ follow the corresponding evaluation of each representation rather than one matched setting.
In particular, the one-hot entry uses $\sigma=1.0$ and the CSSE-CSP entry $\sigma=0.2$, following the projection-time scans in Supplementary Tables~\ref{tab:supp_projection_onehot} and~\ref{tab:supp_projection_csse}.

\begin{table}[htbp]
    \centering
    \caption{\textbf{Metastable and MSUN gain--loss decomposition across element representations.}}
    \label{tab:supp_representation_gainloss}
    \small
    \resizebox{\textwidth}{!}{%
    \begin{tabular}{lrrrrrrrrr}
        \toprule
        Representation &
        Metastable gain &
        Metastable loss &
        MSUN gain &
        MSUN loss &
        $\Delta$Metastable &
        $\Delta$MSUN &
        $A$ &
        $Q$ &
        $Q/A$ \\
        \midrule
        One-hot             &  1.17 & 48.29 &  2.70 & 14.37 & -47.12 & -11.67 & 47.21 &  5.20 & 11.01 \\
        Random Spherical    & 12.45 & 31.71 &  9.85 & 12.48 & -19.26 &  -2.63 & 14.69 &  3.18 & 21.64 \\
        Random Permutation  &  7.24 & 48.28 &  8.39 & 16.80 & -41.04 &  -8.41 & 34.75 &  4.62 & 13.30 \\
        CrystalDiT-style    &  7.72 & 30.81 & 23.80 & 14.35 & -23.09 &   9.45 & 49.55 & 15.94 & 32.16 \\
        Crystalite-style    & 12.08 & 27.73 & 22.15 & 12.49 & -15.65 &   9.66 & 22.66 & 10.34 & 45.64 \\
        CSSE-MLIP           & 12.01 & 18.40 & 23.03 & 12.60 &  -6.39 &  10.43 & 23.94 & 12.46 & 52.06 \\
        CSSE-CSP            &  9.87 & 25.99 & 35.00 & 10.33 & -16.12 &  24.67 & 48.81 & 20.21 & 41.41 \\
        \bottomrule
    \end{tabular}%
    }
\end{table}

\FloatBarrier

\subsection{Evolution of intermediate element assignments with training}
\label{sec:supp_assignments_training}

\begin{figure}[htbp]
    \centering
    \includegraphics[width=\textwidth]{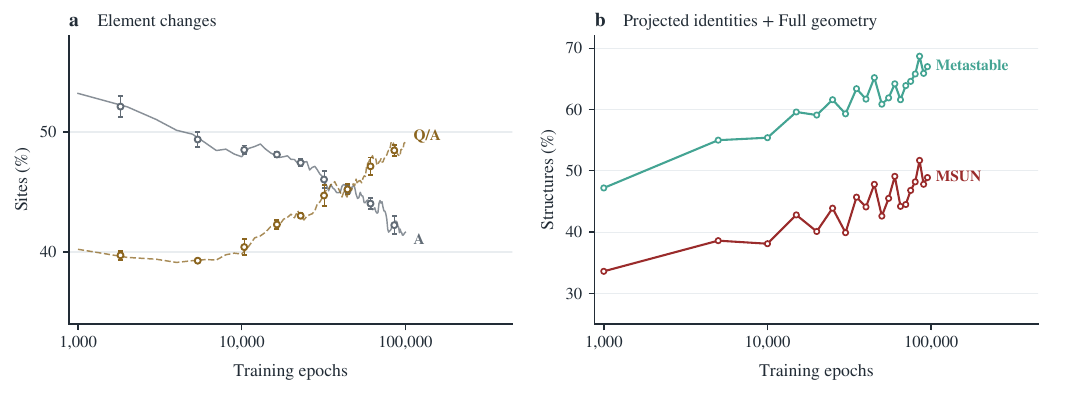}
    \caption{\textbf{Evolution of intermediate element assignments during continued training (DiffCSP, MP20, CSSE-CSP).}
    \textbf{a,} Fraction of all sites whose element identity at the intermediate type state differs from that of the final frame of the same trajectory, $A$, and $Q/A$, the compatible fraction among the differing sites (dashed).
    \textbf{b,} MSUN and Metastable of Projected identities + Full geometry, obtained by assigning the identities projected at $t_{\mathrm{s}}=0.2$ to the terminal coordinates and lattice of the corresponding Full trajectory before relaxation.
    Thin curves in \textbf{a} are moving averages of the 100 checkpoints over nine consecutive checkpoints; large open markers with error bars are the mean and standard deviation over groups of checkpoints that are equally spaced on the logarithmic axis.
    Markers in \textbf{b} are the 20 checkpoints at which structures were relaxed and evaluated.}
    \label{fig:supp_assignments_training}
\end{figure}

Supplementary Fig.~\ref{fig:supp_assignments_training} evaluates the intermediate type state at $t_{\mathrm{s}}=0.2$ against the final frame of the same trajectory, at 100 checkpoints spread over 1,000 to 100,000 epochs of the long training run.
The changed fraction $A$, the compatible fraction $Q$ and their ratio $Q/A$ follow the definitions of Methods.

$A$ decreases from $53\%$ to $42\%$ over the scan, whereas $Q/A$ increases from $40\%$ to $49\%$ (Supplementary Fig.~\ref{fig:supp_assignments_training}a).
Because $Q=A\,(Q/A)$ on the all-site denominator, $Q$ stays within 19--21\% throughout: the fraction of all sites undergoing compatible identity changes remains nearly constant, while their share among all identity changes grows.
Continued training therefore makes the intermediate assignments more conservative, and the changes that remain are more often chemically compatible.

\FloatBarrier

Over the same epochs the structures built from these assignments become progressively more viable.
For Projected identities + Full geometry, Metastable increases from $47\%$ to $67\%$ and MSUN from $34\%$ to $49\%$ (Supplementary Fig.~\ref{fig:supp_assignments_training}b).
This trend is consistent with the sustained MSUN gain of IILS at long training times reported in Fig.~1: as training proceeds, the identities projected at $t_{\mathrm{s}}$ increasingly form structures that survive relaxation under the corresponding Full geometries.

\subsection{Representative terminal structures}
\label{sec:supp_structure_pairs}

\begin{figure}[htbp]
    \centering
    \includegraphics[width=\textwidth]{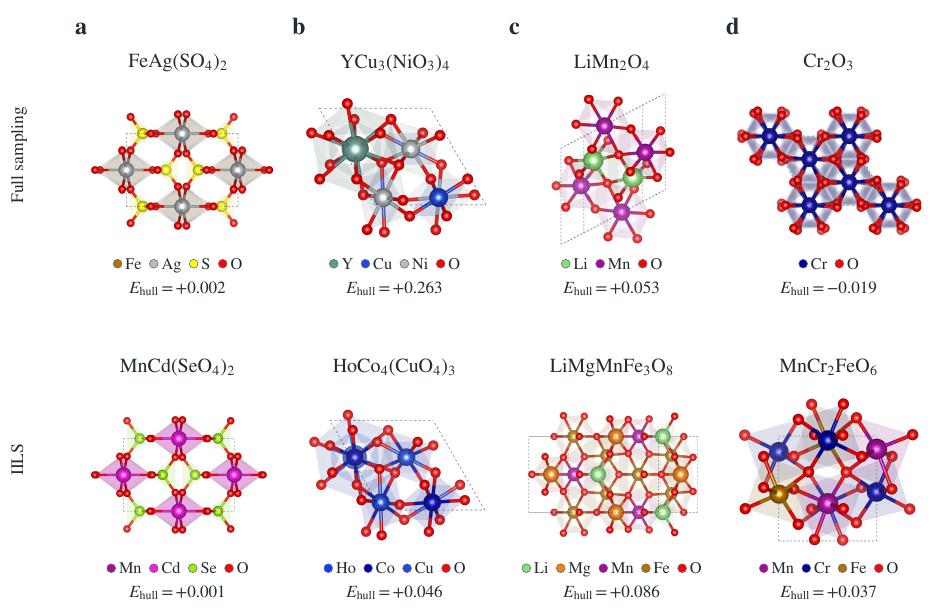}
    \caption{\textbf{Representative terminal structures from shared initial noise.}
    \textbf{a}--\textbf{d}, Four pairs of terminal structures.
    Each column shows the structures obtained from the same initial noise under full sampling (upper) and IILS (lower), with the composition printed above each render and the energy above the convex hull below it.
    The element colours are identified in the legend under every structure.}
    \label{fig:supp_structure_pairs}
\end{figure}

Supplementary Fig.~\ref{fig:supp_structure_pairs} shows four pairs of terminal structures obtained from the same initial noise, with full sampling in the upper row and IILS in the lower row.
The pairs are illustrative and do not enter any of the statistics above; aggregate yields over entire sampling runs are reported in Fig.~1 of the main text.

\subsection{Sampling efficiency and computational cost}
\label{sec:supp_sampling_efficiency}

Generation time is measured for DiffCSP with CSSE-CSP on MP20 using a single NVIDIA A800 GPU.
Each entry of Supplementary Table~\ref{tab:supp_efficiency} is averaged over three repetitions of four batches, with a batch size of 64 and 1,000 reverse-diffusion steps.

\begin{table}[htbp]
    \centering
    \caption{\textbf{Generation cost of full sampling and IILS for CSSE-CSP on MP20.}
    Generation only, on a single A800 GPU with 1,000 reverse steps, batch size 64 and shared initial noise; relaxation and evaluation excluded.
    Throughput is the number of generated structures per second of generation time, and peak memory is the maximum GPU memory allocated during generation.}
    \label{tab:supp_efficiency}
    \small
    \begin{tabular}{lrrr}
        \toprule
        Sampling &
        Time per structure (ms) &
        Throughput (structures/s) &
        Peak memory (MB) \\
        \midrule
        Full & 443 & 2.26 & 298 \\
        IILS & 442 & 2.27 & 294 \\
        \bottomrule
    \end{tabular}
\end{table}

\section{Additional generation evaluation results}
\label{sec:supp_evaluation}

\subsection{Energetics, distributional distances and diversity}
\label{sec:supp_evaluation_metrics}

Supplementary Table~\ref{tab:supp_distribution_diversity} reports the mean formation energy, energy above hull and RMSD of the generated structures, together with additional distributional distance and diversity metrics evaluated using LeMat-GenBench.
The Distribution metrics include Jensen--Shannon (JS) distance, maximum mean discrepancy (MMD) and Fréchet distance (FID), which quantify differences between generated and reference material distributions.
The Diversity metrics characterise variation in elemental compositions (Element), space groups (Space group), atomic sites (Atomic site) and crystal sizes (Crystal size).
All metrics follow the definitions and implementations provided by LeMat-GenBench~\cite{betala2025lemat}.

\begin{table*}[t]
\centering
\begin{threeparttable}
\caption{\textbf{Energetics, distributional distances and diversity across generative models and sampling strategies.}}
\label{tab:supp_distribution_diversity}
\scriptsize
\setlength{\tabcolsep}{4pt}
\renewcommand{\arraystretch}{1.08}
\begin{tabular}{lrrrrrrrrrr}
\toprule
\multirow{2}{*}{Model / Setting} &
\multicolumn{3}{c}{Energetics} &
\multicolumn{3}{c}{Distribution} &
\multicolumn{4}{c}{Diversity} \\
\cmidrule(lr){2-4}\cmidrule(lr){5-7}\cmidrule(lr){8-11}
& $E_{\mathrm{form}}\downarrow$ & $E_{\mathrm{hull}}\downarrow$ & RMSD$\downarrow$ & JS$\downarrow$ & MMD$\downarrow$ & FID$\downarrow$ & ElemDiv$\uparrow$ & SGDiv$\uparrow$ & SiteDiv$\uparrow$ & SizeDiv$\uparrow$ \\
\midrule
\multicolumn{11}{l}{\textbf{Previous methods}} \\
\addlinespace[1pt]
Creator-MV-DiT & -0.94 & \textbf{0.06} & 0.16 & 0.35 & 0.004 & 1.95 & 0.71 & 0.37 & 15.02 & 0.25 \\
Crystalite & -0.89 & 0.09 & 0.13 &  0.28 &  0.006 &  1.84 &  0.72 &  0.46 &  14.48 & 0.24 \\
Chemeleon2 & -0.72 & 0.15 & 0.41 &  0.55 &  0.009 &  2.30 &  0.67 &  0.03 &  13.85 & 0.23 \\
MatterGen & -0.71 & 0.17 & 0.39 &  0.44 &  0.006 &  2.13 &  0.64 &  0.14 &  12.78 & 0.24 \\
OMatG-FC & -0.92 & \underline{0.07} & \textbf{0.06} & 0.26 & 0.004 & 2.08 & 0.74 & 0.57 & 14.31 & 0.25 \\
ADiT & -0.74 & 0.33 & 0.38 & 0.51 & 0.002 & 1.72 & 0.71 & 0.03 & 14.18 & 0.24 \\
PLaID++ & -0.50 & 0.08 & 0.13 & 0.43 & 0.029 & 2.69 & 0.68 & 0.26 & 6.01 & 0.21 \\
MiAD & -0.85 & 0.08 & 0.24 & 0.48 & 0.002 & 1.92 & 0.72 & 0.07 & 12.40 & -- \\

\midrule
\multicolumn{11}{l}{\textbf{DiffCSP-based models}} \\
\addlinespace[1pt]
One-hot & -0.73 & 0.12 & 0.23 & 0.35 & 0.006 & 1.59 & 0.64 & 0.33 & 14.41 & 0.24 \\
\quad + IILS & -0.42 & 0.40 & 0.45 & 0.33 & 0.002 & 0.97 & 0.64 & 0.18 & 10.41 & 0.26 \\
Crystalite-style & -0.87 & 0.08 & 0.18 & 0.35 & 0.006 & 1.86 & 0.64 & 0.32 & 14.56 & 0.24 \\
\quad + IILS & -0.90 & 0.12 & 0.21 & 0.41 & 0.013 & 2.96 & 0.64 & 0.30 & 14.29 & 0.24 \\
CrystalDiT-style & -0.68 & 0.08 & 0.17 & 0.38 & 0.002 & 1.96 & 0.64 & 0.30 & 14.59 & 0.24 \\
\quad + IILS & -0.27 & 0.15 & 0.22 & 0.46 & 0.014 & 2.71 & 0.64 & 0.26 & 14.61 & 0.24 \\
CSSE-MLIP & -0.89 & 0.09 & 0.21 & 0.38 & 0.011 & 1.99 & 0.64 & 0.31 & 14.52 & 0.25 \\
\quad + IILS & -0.90 & 0.08 & 0.24 & 0.40 & 0.018 & 2.88 & 0.64 & 0.27 & 14.23 & 0.25 \\
CSSE-CSP & -0.84 & 0.08 & 0.21 & 0.36 & 0.009 & 1.81 & 0.64 & 0.31 & 14.52 & 0.25 \\
\quad + IILS & -0.73 & 0.11 & 0.18 & 0.41 & 0.006 & 2.39 & 0.64 & 0.23 & 14.86 & 0.24 \\
\quad + IILS (LT) & -0.76 & 0.08 & 0.14 & 0.39 & 0.005 & 2.15 & 0.64 & 0.27 & 15.26 & 0.24 \\

\midrule
\multicolumn{11}{l}{\textbf{Crystalite-based models}} \\
\addlinespace[1pt]
Original representation & -0.93 & 0.10 & 0.20 & 0.40 & 0.008 & 1.87 & 0.65 & 0.28 & 14.78 & 0.24  \\
\quad + IILS & -0.95 & 0.08 & 0.14 & 0.40 & 0.007 & 1.99 & 0.66 & 0.30 & 14.71 & 0.24 \\
CSSE-CSP & -0.90 & 0.09 & 0.20 & 0.39 & 0.008 & 1.82 & 0.65 & 0.28 & 14.68 & 0.24 \\
\quad + IILS & -0.93 & 0.08 & 0.14 & 0.39 & 0.008 & 2.04 & 0.65 & 0.28 & 14.83 & 0.24 \\
\quad + IILS (LT) & \underline{-0.96} & \textbf{0.06} & \underline{0.11} & 0.35 & 0.007 & 2.14 & 0.64 & 0.35 & 14.82 & 0.24 \\

\midrule
\multicolumn{11}{l}{\textbf{AlexMP20 DiffCSP-based}} \\
\addlinespace[1pt]
One-hot & -0.56 & 0.08 & 0.19 & 0.30 & 0.001 & 1.18 & 0.64 & 0.33 & 14.27 & 0.25 \\
\quad + IILS & -0.52 & 0.12 & 0.31 & 0.38 & 0.014 & 3.01 & 0.64 & 0.27 & 14.41 & 0.25 \\
CSSE-CSP & -0.60 & 0.06 & 0.16 & 0.32 & 0.001 & 1.23 & 0.64 & 0.29 & 14.24 & 0.25 \\
\quad + IILS & -0.49 & 0.06 & 0.18 & 0.44 & 0.012 & 2.62 & 0.63 & 0.22 & 15.19 & 0.24 \\

\midrule
\multicolumn{11}{l}{\textbf{Terminal element-substitution controls (DiffCSP, CSSE-CSP)}} \\
\addlinespace[1pt]
Codebook substitution & -0.24 & 0.48 & 0.51 & 0.32 & 0.006 & 1.70 & 0.64 & 0.11 & 8.32 & 0.28 \\
Hautier substitution & -0.46 & 0.33 & 0.41 & 0.40 & 0.008 & 1.98 & 0.64 & 0.13 & 13.53 & 0.28 \\
\midrule

\multicolumn{11}{l}{
\textbf{Template-based methods}
} \\
\addlinespace[1pt]
Ion exchange & \textbf{-1.45} & 0.10 & 0.17 & 0.35 & 0.076 & 4.80 & 0.64 & 0.33 & 13.73 & 0.24 \\

\bottomrule
\end{tabular}
\begin{tablenotes}[flushleft]
\footnotesize
\item Energetics are reported as means after relaxation with MatterSim-1M~\cite{yang2024mattersim}; energies are in eV\,atom$^{-1}$ and RMSD in \AA{}.
Bold and underlined values in the Energetics columns mark the best and second-best results, respectively, among the settings evaluated on MP20; ties are all marked.
\end{tablenotes}
\end{threeparttable}
\end{table*}

\FloatBarrier

\subsection{Composition-count distribution and per-group discovery performance}
\label{sec:supp_composition_counts}

\begin{figure}[htbp]
    \centering
    \includegraphics[width=0.85\textwidth]{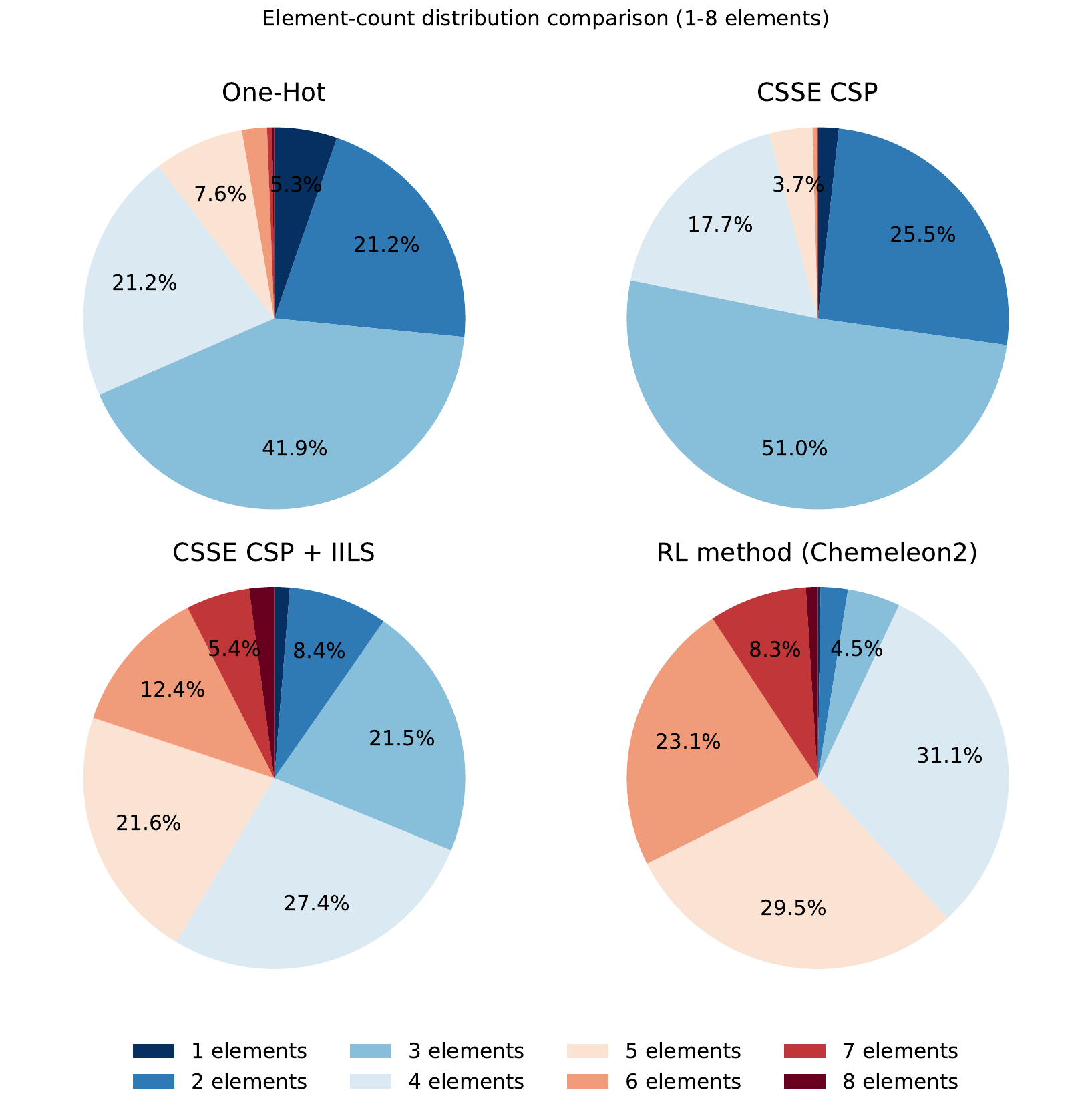}
    \caption{\textbf{Composition-count distribution across sampling strategies and previous methods.}
    Normalised distributions of the number of distinct element types per generated structure for full sampling and IILS, compared with previously published generative models.}
    \label{fig:supp_composition_count}
\end{figure}

For each generated structure we record the number of distinct element types in its composition, and compare its distribution across sampling strategies.
Supplementary Fig.~\ref{fig:supp_composition_count} compares the distributions of the number of distinct elements per generated structure for DiffCSP with the CSSE-CSP representation under full sampling and under IILS, and additionally reports the corresponding distributions of the original DiffCSP model and of a frontier RL-based generative model.

The chemically aligned CSSE-CSP codebook lowers the mean composition count relative to the one-hot codebook (2.96 versus 3.14 distinct elements).
Under IILS the mean composition count rises to 4.28, and 41.1\% of the structures contain five or more elements, far above the 3.6\% obtained under full sampling with the same representation.
This fraction is lower than that of the RL-based model (61.9\%).
Despite its more moderate composition count, IILS reaches an MSUN yield approximately twice that of the RL-based model (Table~1 of the main text), indicating that the discovery gain is not obtained by simply driving the composition distribution towards higher element counts.

To distinguish the origin of the MSUN gain, Supplementary Table~\ref{tab:supp_msun_by_composition} compares the two sampling strategies within each composition group.
IILS yields a higher MSUN fraction in every group with three or more elements, whereas the binary group is lower under IILS (20.0\% versus 22.2\%).
Restricting the composition space to at most four distinct elements preserves the advantage of IILS over full sampling (Supplementary Table~\ref{tab:supp_comp_element_controls}).
The MSUN advantage therefore cannot be explained solely by the shift towards compositions containing more distinct elements.

\begin{table}[htbp]
    \centering
    \caption{\textbf{Composition mix and discovery performance by composition count under full sampling and IILS.}
    Structures are grouped by the number of distinct elements, for DiffCSP on MP20 with the CSSE-CSP representation.
    Share is the fraction of generated structures in the group, and $\Delta$MSUN is IILS minus full sampling.
    The remaining share corresponds to unary and $\ge$7-element compositions.}
    \label{tab:supp_msun_by_composition}
    \small
    \setlength{\tabcolsep}{4pt}
    \resizebox{\textwidth}{!}{%
    \begin{tabular}{lrrrrrrr}
        \toprule
        \multirow{2}{*}{Composition} &
        \multicolumn{3}{c}{Full sampling} &
        \multicolumn{3}{c}{IILS ($t_s=0.2$, $\sigma=0.2$)} &
        \multirow{2}{*}{$\Delta$MSUN (pp)} \\
        \cmidrule(lr){2-4}\cmidrule(lr){5-7}
        & Share (\%) & Metastable (\%) & MSUN (\%) & Share (\%) & Metastable (\%) & MSUN (\%) & \\
        \midrule
        Binary     & 25.7 & 78.5 & 22.2 & 8.5  & 79.2 & 20.0 & $-2.2$ \\
        Ternary    & 51.7 & 74.3 & 20.0 & 21.3 & 72.7 & 31.7 & $+11.7$ \\
        Quaternary & 17.3 & 57.4 & 28.6 & 27.8 & 62.4 & 46.8 & $+18.2$ \\
        Quinary    & 3.2  & 45.2 & 32.8 & 21.5 & 53.8 & 52.7 & $+19.9$ \\
        Hexary     & 0.4  & 22.7 & 22.7 & 12.3 & 48.4 & 48.4 & $+25.6$ \\
        \bottomrule
    \end{tabular}%
    }
\end{table}

\FloatBarrier

\subsection{Element occurrence frequencies}
\label{sec:supp_element_frequencies}

\begin{figure}[htbp]
    \centering
    \includegraphics[width=\textwidth]{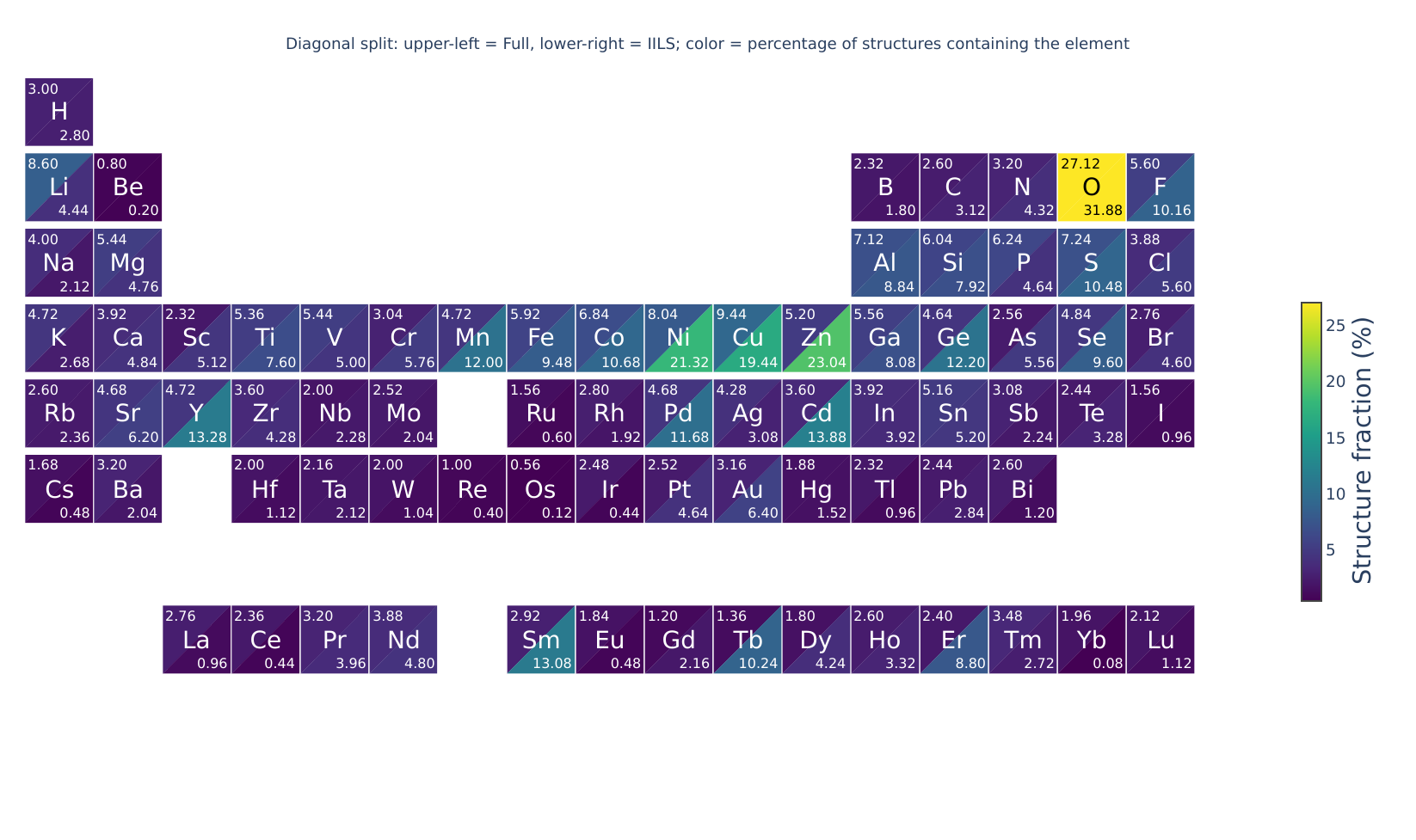}
    \caption{\textbf{Element occurrence frequencies under full sampling and IILS.}
    Fraction of generated structures whose composition contains each element, for DiffCSP with the CSSE-CSP representation under full sampling and under IILS ($t_s=0.2$, $\sigma=0.2$), over all elements present in the generated structures.}
    \label{fig:supp_element_occurrence}
\end{figure}

\begin{figure}[htbp]
    \centering
    \includegraphics[width=\textwidth]{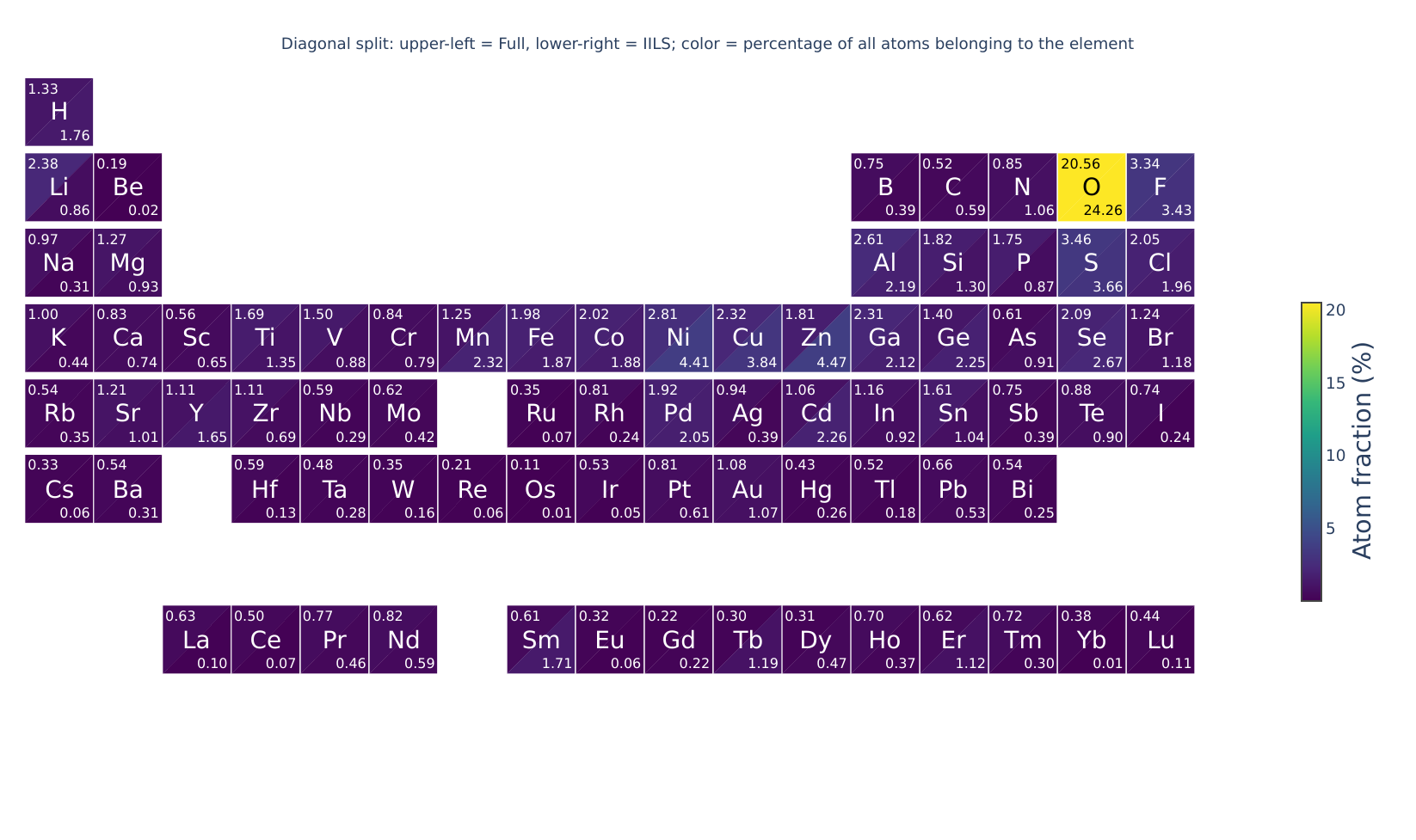}
    \caption{\textbf{Atom-weighted element occurrence frequencies under full sampling and IILS.}
    Fraction of all atoms that belong to each element, for DiffCSP with the CSSE-CSP representation under full sampling and under IILS ($t_s=0.2$, $\sigma=0.2$), over all elements present in the generated structures.
    Here each element is weighted by the number of sites it occupies, whereas Supplementary Fig.~\ref{fig:supp_element_occurrence} weights every structure equally.}
    \label{fig:supp_element_atom_fraction}
\end{figure}

\begin{figure}[htbp]
    \centering
    \includegraphics[width=\textwidth]{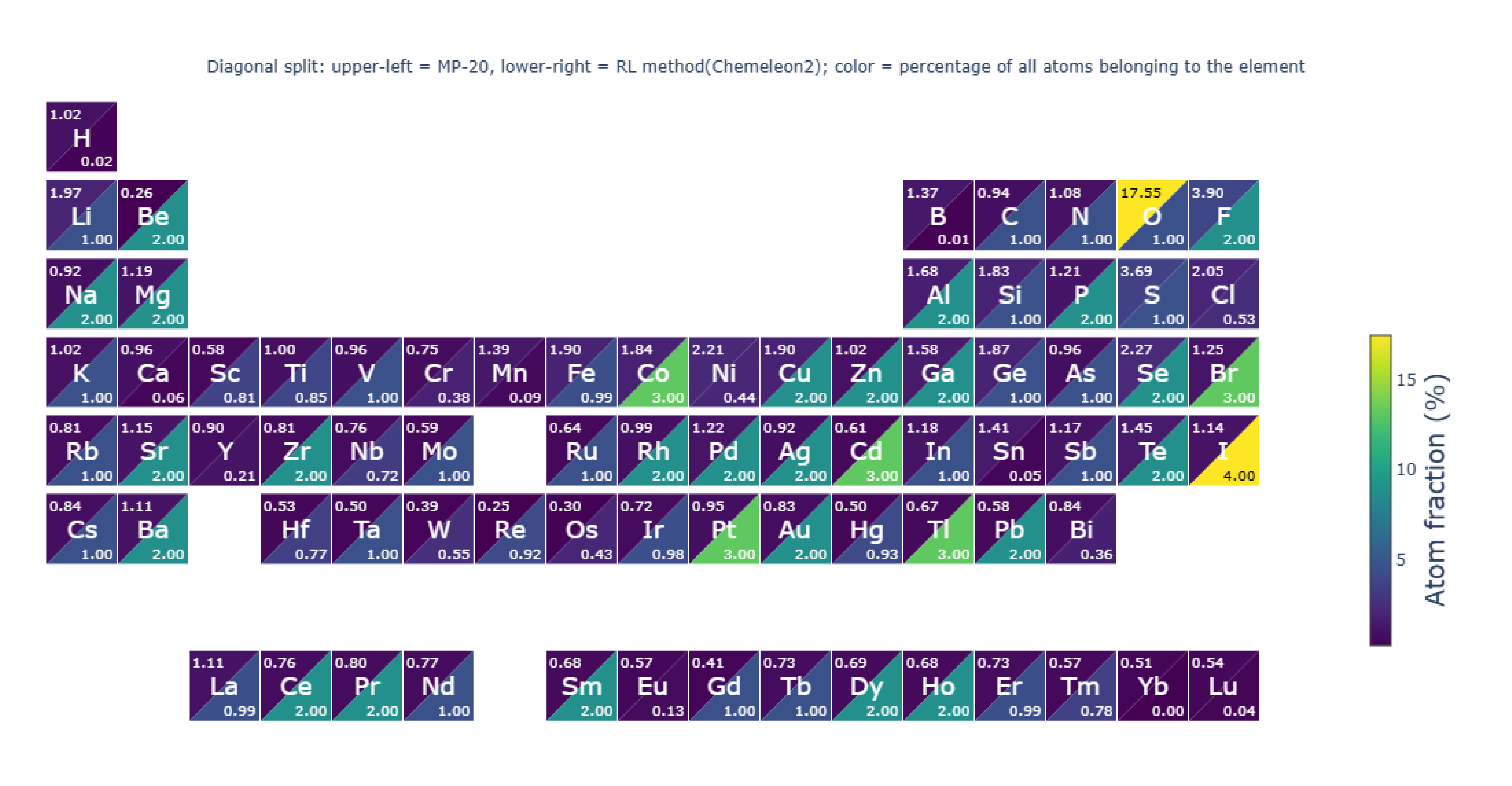}
    \caption{\textbf{Atom-weighted element frequencies in the MP20 training set and Reinforcement Learning based Model.}
    Fraction of all atoms that belong to each element, for MP-20 training set and reinforcement learning based model(Chemeleon2), over all elements present in the generated structures.
    The data of Chemeleon2 is derived from its official paper~\cite{park2026chemeleon2}.}
    \label{fig:supp_element_atom_fraction_mp20_rl}
\end{figure}

Supplementary Figs.~\ref{fig:supp_element_occurrence} and~\ref{fig:supp_element_atom_fraction} compare element frequencies under Full and IILS, measured by the fraction of generated structures containing each element and by its fraction of all atomic sites, respectively. IILS increases the occurrence of certain elements, including Zn, Cu, Sm and Y, across generated structures, while their atom-weighted frequencies change less markedly. Compared with the MP20 training distribution, IILS exhibits a relatively moderate shift in atom-weighted element frequencies, particularly in comparison with the more pronounced redistribution observed for the RL-based Chemeleon2 model (Supplementary Fig.~\ref{fig:supp_element_atom_fraction_mp20_rl}). The MSUN advantage of IILS also persists after excluding rare-earth-containing structures (Supplementary Table~\ref{tab:supp_comp_element_controls}).

\begin{table}[htbp]
    \centering
    \caption{\textbf{Controls for composition count and element concentration.}
    DiffCSP with the CSSE-CSP representation on MP20.
    In each block the composition space is restricted either by excluding rare-earth elements or by excluding compositions with more than four distinct elements, and the two sampling strategies are compared under the same restriction.
    The reference values are those of Table~1 of the main text, averaged over three sampling runs; the restricted settings are evaluated on a single run, and the restricted IILS settings retain fewer valid structures.
    Metastable includes Stable, and MSUN includes SUN.}
    \label{tab:supp_comp_element_controls}
    \scriptsize
    \setlength{\tabcolsep}{3.5pt}
    \renewcommand{\arraystretch}{1.08}
    \begin{tabular}{l ccc ccc cc cc}
        \toprule
        \multirow{2}{*}{Setting}
        & \multirow{2}{*}{Valid$\uparrow$}
        & \multirow{2}{*}{Unique$\uparrow$}
        & \multirow{2}{*}{Novel$\uparrow$}
        & \multicolumn{3}{c}{Energetics}
        & \multicolumn{2}{c}{Stability}
        & \multicolumn{2}{c}{Metastability} \\
        \cmidrule(lr){5-7}\cmidrule(lr){8-9}\cmidrule(lr){10-11}
        & & & & $E_{\mathrm{form}}\downarrow$ & $E_{\mathrm{hull}}\downarrow$ & RMSD$\downarrow$
        & Stable$\uparrow$ & SUN$\uparrow$ & Metastable$\uparrow$ & MSUN$\uparrow$ \\
        \midrule
        \multicolumn{11}{l}{\textbf{No restriction}} \\
        \addlinespace[1pt]
        Full sampling          & 97.5 & 95.9 & 51.9 & -0.84 & 0.08 & 0.21 & 4.0 & 0.7 & 67.6 & 24.9 \\
        IILS                   & 98.9 & 95.9 & 77.9 & -0.73 & 0.11 & 0.18 & 1.9 & 1.0 & 61.8 & 42.4 \\
        \midrule
        \multicolumn{11}{l}{\textbf{Excluding rare-earth elements}} \\
        \addlinespace[1pt]
        Full sampling          & 97.6 & 93.7 & 54.6 & -0.84 & 0.09 & 0.19 & 3.0 & 0.2 & 64.8 & 24.9 \\
        IILS                   & 98.6 & 96.1 & 80.0 & -0.71 & 0.14 & 0.24 & 1.0 & 0.7 & 49.0 & 32.8 \\
        \midrule
        \multicolumn{11}{l}{\textbf{At most four distinct elements}} \\
        \addlinespace[1pt]
        Full sampling          & 98.1 & 95.5 & 43.8 & -0.84 & 0.07 & 0.15 & 4.5 & 0.5 & 72.8 & 21.5 \\
        IILS                   & 98.5 & 92.5 & 62.0 & -0.66 & 0.09 & 0.16 & 2.9 & 1.2 & 68.8 & 35.2 \\
        \midrule
        \multicolumn{11}{l}{\textbf{Both restrictions}} \\
        \addlinespace[1pt]
        Full sampling          & 97.4 & 93.3 & 52.8 & -0.81 & 0.09 & 0.19 & 3.1 & 0.2 & 66.2 & 24.7 \\
        IILS                   & 97.8 & 92.9 & 69.7 & -0.64 & 0.12 & 0.21 & 1.5 & 0.8 & 58.6 & 33.9 \\
        \bottomrule
    \end{tabular}
\end{table}

\FloatBarrier

\subsection{Symmetry-defined prototype novelty}
\label{sec:supp_prototype_novelty}

The Novel and MSUN metrics in the main text are based on full-structure matching against a reference set and therefore count a generated crystal as novel whenever it differs from any reference structure, including the case of a pure element substitution on an existing structural framework.
To assess novelty beyond elemental substitutions, we additionally characterize each generated structure by a symmetry-defined prototype and determine whether it is present in the MP20 training split.

For a generated or reference structure, we determine the space group and the symmetry-distinct Wyckoff positions with \texttt{pymatgen}'s \texttt{SpacegroupAnalyzer} (symmetry tolerance $\mathrm{symprec}=0.1$~\AA).
The prototype is defined as the pair
\begin{equation}
\Pi = \bigl(\mathrm{SG},\,\{\,m_j\,w_j\,\}_j\bigr),
\label{eq:supp_prototype}
\end{equation}
where $\mathrm{SG}$ is the space-group number and $\{m_j w_j\}_j$ is the multiset of occupied Wyckoff positions, each written as its multiplicity $m_j$ and Wyckoff letter $w_j$ (for example, $\mathrm{SG}=225$ with $\{4a,8c\}$).
The prototype encodes the space-group symmetry and occupied Wyckoff positions while remaining invariant to the chemical identities of the occupying elements, so two structures that differ only by an element substitution on the same framework share the same prototype.
The converse does not hold: structures sharing a prototype may still differ in lattice parameters, cell shape, or Wyckoff coordinates within the same sites.

We extract the prototype $\Pi$ of every structure in the MP20 training split and collect the set $\mathcal{P}_{\mathrm{MP20}}$ of distinct prototypes.
A generated structure is counted as having an \emph{unseen prototype} if its prototype is not in $\mathcal{P}_{\mathrm{MP20}}$, and otherwise as sharing the symmetry descriptors of an existing MP20 framework.
The symmetry-defined prototype novelty is the fraction of structures whose prototype is unseen in MP20,
\begin{equation}
\mathrm{PNov} = \frac{\bigl|\{\,s : \Pi(s)\notin\mathcal{P}_{\mathrm{MP20}}\,\}\bigr|}{N},
\label{eq:supp_pnov}
\end{equation}
where $N$ is the number of structures in the set over which the fraction is taken.
We report $\mathrm{PNov}$ separately over all generated structures (All) and over the MSUN set, using the same generated samples and validity filtering as the main benchmark.
Supplementary Table~\ref{tab:supp_prototype_novelty} reports $\mathrm{PNov}$ across models and sampling strategies.

\begin{table}[htbp]
    \centering
    \caption{\textbf{Symmetry-defined prototype novelty against MP20.}
    Fraction of generated structures whose prototype (space group and occupied Wyckoff positions, Eq.~\eqref{eq:supp_prototype}) is absent from the MP20 training split; the complementary fraction corresponds to structures sharing the symmetry descriptors of an existing MP20 framework.
    $\mathrm{PNov}$ is reported over the MSUN set and over all generated structures (All).
    MSUN denotes the fraction satisfying metastability, uniqueness and novelty, as defined in the main text.}
    \label{tab:supp_prototype_novelty}
    \small
    \begin{tabular}{llrr}
        \toprule
        Base model & Representation / Setting &
        PNov$_{\mathrm{MSUN}}$ (\%) & PNov$_{\mathrm{All}}$ (\%) \\
        \midrule
        \multirow{3}{*}{DiffCSP}
          & One-hot            & 23.5 & 23.5 \\
          & CSSE-CSP           & 29.2 & 19.3 \\
          & CSSE-CSP + IILS    & 27.8 & 21.9 \\
        \midrule
        \multirow{4}{*}{Crystalite}
          & Original representation         & 18.1 & 10.7 \\
          & Original representation + IILS  & 21.2 & 13.8 \\
          & CSSE-CSP                        & 16.4 & 11.5 \\
          & CSSE-CSP + IILS                 & 17.9 & 13.4 \\
        \bottomrule
    \end{tabular}
\end{table}

\FloatBarrier

CSSE-CSP increases $\mathrm{PNov}_{\mathrm{MSUN}}$ relative to one-hot encoding for DiffCSP (23.5\% to 29.2\%). 
The effect of IILS varies across architectures: $\mathrm{PNov}_{\mathrm{MSUN}}$ decreases slightly under DiffCSP (29.2\% to 27.8\%) but increases under Crystalite (16.4\% to 17.9\%).

\subsection{Space group and crystal system distributions}
\label{sec:supp_space_groups}

\begin{figure}[htbp]
    \centering
    \includegraphics[width=\textwidth]{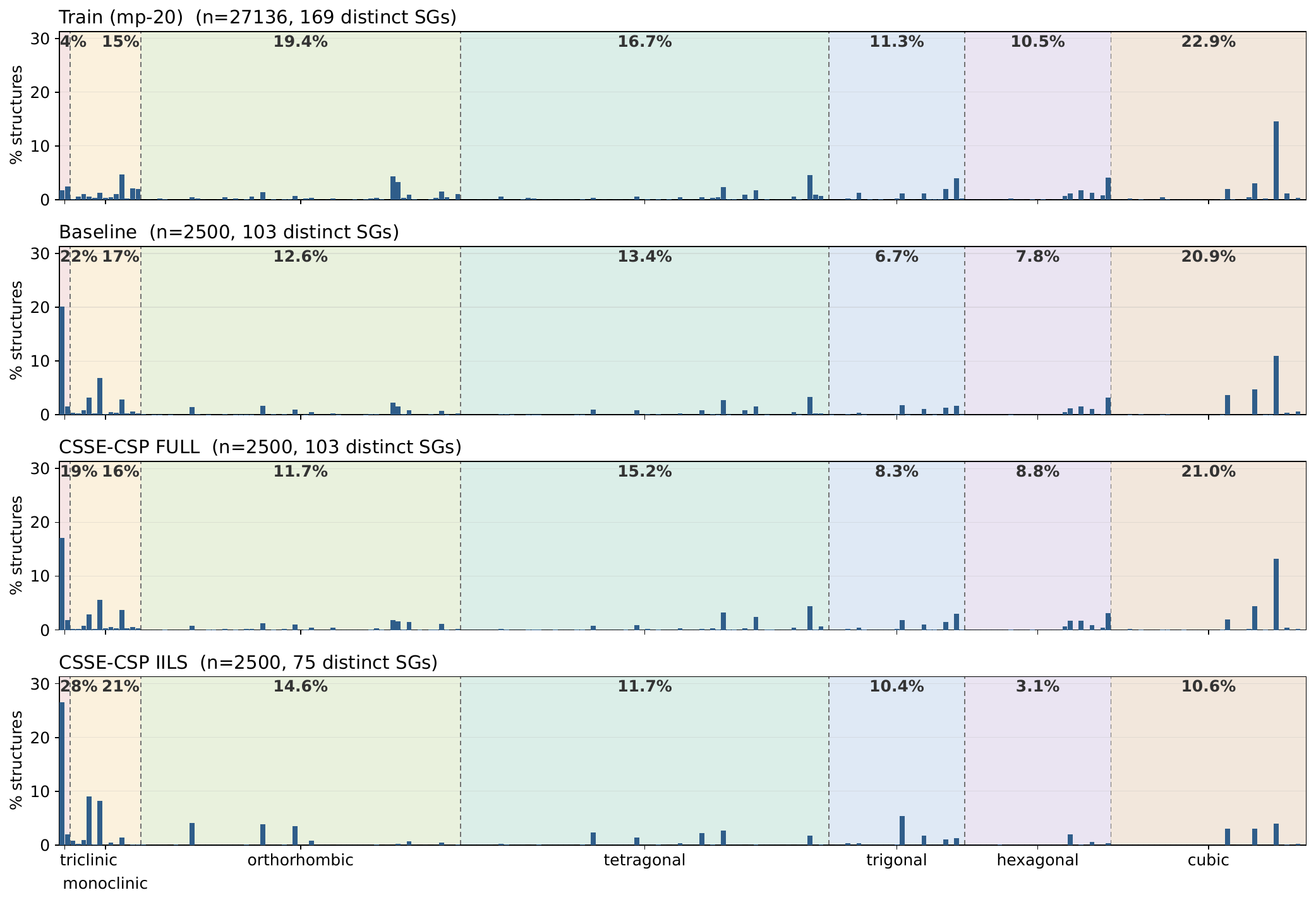}
    \caption{\textbf{Space group and crystal system distributions.}
    Fraction of structures assigned to each of the 230 space groups, for the MP20 training split and for generated structures from the one-hot baseline under full sampling and from CSSE-CSP under full sampling and under IILS ($t_s=0.2$, $\sigma=0.2$), from top to bottom.
    Space groups are assigned with \texttt{pymatgen}'s \texttt{SpacegroupAnalyzer} ($\mathrm{symprec}=0.1$~\AA) to the relaxed structures.
    Shaded bands and dashed separators mark the seven crystal systems, and the fraction of structures in each is annotated inside the band.}
    \label{fig:supp_spacegroup}
\end{figure}

IILS shifts the space-group and crystal-system distributions of the generated structures towards lower symmetry (Supplementary Fig.~\ref{fig:supp_spacegroup}).
All structures entering this comparison are relaxed.
Comparing the sample pairs shared by the two trajectories, 65.5\% change space group and 46.9\% change crystal system relative to full sampling under the same configuration, and 92\% of the space-group changes lower the symmetry.
Most pairs nevertheless keep the same structural backbone: with \texttt{pymatgen}'s \texttt{StructureMatcher} ($\mathrm{ltol}=0.3$, $\mathrm{stol}=0.5$, $\mathrm{angle\_tol}=8^\circ$, all other settings at their default values) with element identities discarded, 75.2\% of the pairs match, rising to 80.1\% when the second structure is placed in the cell of the first so that only the fractional coordinates are compared.

\subsection{Representation-space projection of generated structures}
\label{sec:supp_tsne}

Each crystal is represented by the mean of its 128-dimensional per-atom features extracted using a frozen DPA3 3.1M encoder~\cite{zhang2026dpa3}, which also serves as the source model for CSSE-MLIP (Supplementary Note~\ref{sec:supp_codebook_sources}).
The analysis includes all 27,136 MP20 training structures and one batch of 2,500 relaxed generated structures from each of Full and IILS.
Both sampling methods use the same CSSE-CSP DiffCSP checkpoint, random seed and element-channel noise scale $\sigma=0.2$, with projection times $t_s=0$ and $0.2$, respectively.

The structure representations are concatenated and jointly projected into two dimensions using t-SNE~\cite{vandermaaten2008tsne}, with no feature scaling, Euclidean distance, perplexity 30, PCA initialization.
The two views use identical coordinate limits, and all generated samples are displayed.
The background density contours are obtained from a two-dimensional Gaussian kernel density estimate fitted only to the projected MP20 training structures, using Scott's bandwidth rule.

\subsection{Alternative evaluation pipeline}
\label{sec:supp_alternative_evaluation}
Supplementary Table~\ref{tab:supp_mattergen_evaluation} reports performance under the MatterGen evaluation pipeline~\cite{zeni2025mattergen}, whose Stable and S.U.N. columns correspond to the Metastable and MSUN fractions defined in the main text.

\begin{table*}[t]
\centering
\begin{threeparttable}
\caption{\textbf{Crystal generation and materials discovery performance under the MatterGen evaluation pipeline.}}
\label{tab:supp_mattergen_evaluation}
\scriptsize
\setlength{\tabcolsep}{4.2pt}
\renewcommand{\arraystretch}{1.08}
\begin{tabular}{lrrrrrrrr}
\toprule
\multirow{3}{*}{Model} &
\multicolumn{4}{c}{Validity and Novelty} &
\multicolumn{4}{c}{Stability and Relaxation} \\
\cmidrule(lr){2-5}\cmidrule(lr){6-9}
& Struct. Val. & Comp. Val. & Unique & Novel & Stable & S.U.N. & Avg. Hull & Avg. RMSD \\
& (\%) $\uparrow$ & (\%) $\uparrow$ & (\%) $\uparrow$ & (\%) $\uparrow$
& (\%) $\uparrow$ & (\%) $\uparrow$ & (eV/atom) $\downarrow$ & (\AA) $\downarrow$ \\
\midrule
\multicolumn{9}{l}{\textbf{Previous methods}} \\
\addlinespace[1pt]
MatterGen & 100.0 & 83.5 & 97.9 & 75.0 & 45.7 & 23.8 & 0.182 & 0.15 \\
ADiT & 100.0 & 90.2 & 89.6 & 43.2 & 70.0 & 17.2 & 0.148 & 0.49 \\

\midrule
\multicolumn{9}{l}{\textbf{DiffCSP-based models}} \\
\addlinespace[1pt]
One-hot & 100.0 & 85.5 & 96.9 & 64.7 & 53.7 & 21.6 & 0.167 & 0.28 \\
\quad + IILS & 100.0 & 76.9 & 99.8 & 85.9 & 15.8 & 4.7 & 0.407 & 0.57 \\
Crystalite-style & 100.0 & 85.9 & 97.8 & 52.4 & 67.7 & 23.5 & 0.113 & 0.21 \\
\quad + IILS & 100.0 & 88.5 & 94.1 & 73.1 & 51.1 & 26.9 & 0.190 & 0.33 \\
CrystalDiT-style & 100.0 & 86.7 & 94.9 & 55.7 & 65.5 & 24.2 & 0.136 & 0.27 \\
\quad + IILS & 100.0 & 91.7 & 89.4 & 78.4 & 49.1 & 28.0 & 0.191 & 0.36 \\
CSSE-MLIP & 100.0 & 83.5 & 98.2 & 56.8 & 65.4 & 25.4 & 0.135 & 0.30 \\
\quad + IILS & 100.0 & 87.8 & 90.9 & 60.9 & 70.9 & 34.0 & 0.111 & 0.34 \\
CSSE-CSP & 100.0 & 86.0 & 98.0 & 55.4 & 68.8 & 27.3 & 0.107 & 0.25 \\
\quad + IILS & 100.0 & 90.2 & 96.2 & 78.4 & 59.3 & 39.5 & 0.156 & 0.25 \\
\quad + IILS (LT) & 100.0 & 91.0 & 96.6 & 77.0 & 69.9 & 49.2 & 0.131 & 0.14 \\

\midrule
\multicolumn{9}{l}{\textbf{Crystalite-based models}} \\
\addlinespace[1pt]
Original representation & 100.0 & 89.2 & 96.9 & 27.6 & 81.5 & 14.6 & 0.117 & 0.16 \\
\quad + IILS & 100.0 & 90.8 & 98.8 & 60.3 & 64.7 & 28.8 & 0.172 & 0.18 \\
CSSE-MLIP & 100.0 & 89.2 & 96.6 & 24.2 & 83.9 & 13.5 & 0.103 & 0.13 \\
\quad + IILS & 100.0 & 89.9 & 98.0 & 49.7 & 78.5 & 32.4 & 0.115 & 0.15 \\
CSSE-CSP & 100.0 & 88.7 & 97.3 & 29.6 & 83.2 & 16.9 & 0.103 & 0.15 \\
\quad + IILS & 100.0 & 90.1 & 99.3 & 66.5 & 70.7 & 39.6 & 0.136 & 0.17 \\
\quad + IILS (LT) & 100.0 & 90.7 & 99.2 & 59.2 & 78.1 & 40.8 & 0.117 & 0.07 \\

\midrule
\multicolumn{9}{l}{\textbf{AlexMP20 DiffCSP-based}} \\
\addlinespace[1pt]
One-hot & 100.0 & 88.7 & 99.5 & 63.6 & 70.6 & 37.0 & 0.081 & 0.14 \\
\quad + IILS & 100.0 & 81.1 & 93.0 & 67.9 & 59.0 & 28.7 & 0.126 & 0.35 \\
CSSE-MLIP & 100.0 & 88.2 & 99.7 & 50.9 & 79.9 & 34.4 & 0.063 & 0.16 \\
\quad + IILS & 100.0 & 90.4 & 96.5 & 71.0 & 75.4 & 47.5 & 0.090 & 0.27 \\
CSSE-CSP & 100.0 & 88.7 & 99.8 & 51.5 & 81.0 & 36.3 & 0.060 & 0.12 \\
\quad + IILS & 100.0 & 93.7 & 98.2 & 83.3 & 73.8 & 58.0 & 0.085 & 0.15 \\

\midrule
\multicolumn{9}{l}{\textbf{Terminal element-substitution controls (DiffCSP, CSSE-CSP)}} \\
\addlinespace[1pt]
Codebook substitution & 100.0 & 66.9 & 100.0 & 92.5 & 4.9 & 1.0 & 0.510 & 0.63 \\
\quad + renoise & 100.0 & 72.1 & 100.0 & 92.9 & 6.5 & 2.1 & 0.465 & 0.59 \\
Hautier substitution & 100.0 & 86.6 & 99.3 & 92.8 & 20.8 & 15.1 & 0.311 & 0.49 \\
\quad + renoise & 100.0 & 88.7 & 99.9 & 91.6 & 19.6 & 13.2 & 0.324 & 0.49 \\
\bottomrule
\end{tabular}
\end{threeparttable}
\end{table*}

\FloatBarrier

\subsection{Continued-training trends on additional architectures and datasets}
\label{sec:supp_continued_training}

\begin{figure}[htbp]
    \centering
    \includegraphics[width=\textwidth]{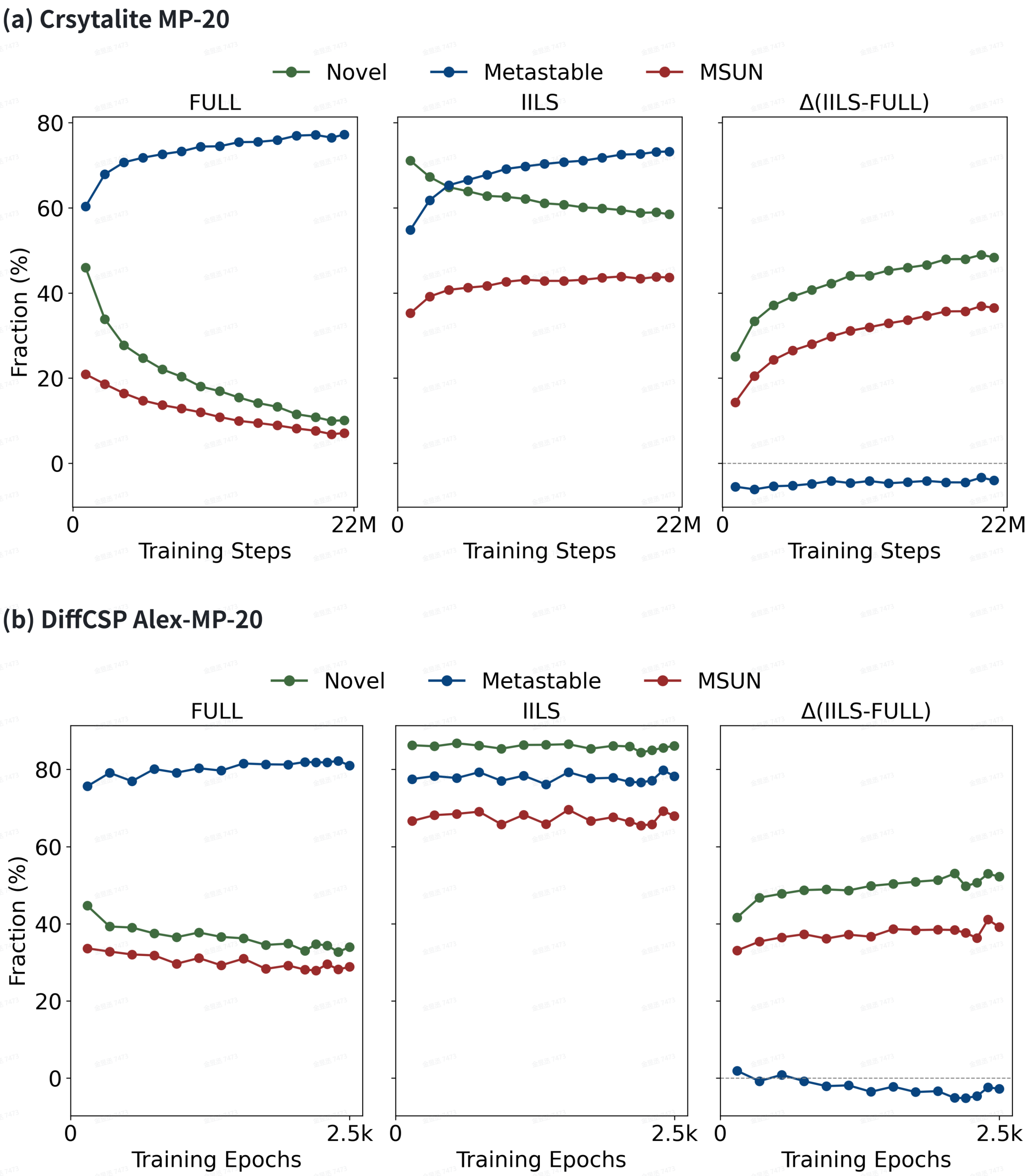}
    \caption{\textbf{Continued-training trends on additional architectures and datasets.}
    \textbf{a,} Evolution of Novel, Metastable and MSUN under IILS and full sampling, together with their differences, during continued training of Crystalite with CSSE-CSP on MP20.
    \textbf{b,} Same as \textbf{a}, for DiffCSP with CSSE-CSP on AlexMP20.}
    \label{fig:supp_continued_training}
\end{figure}

Supplementary Fig.~\ref{fig:supp_continued_training} extends the training comparison in Fig.~1 to Crystalite on MP20 and DiffCSP on AlexMP20. For Crystalite, Full MSUN declines after an early maximum, whereas IILS continues to improve with training. On AlexMP20, Full MSUN also declines after its early maximum, while IILS maintains a high MSUN level without comparable degradation. These results show that the divergent training trends are not restricted to a single architecture or dataset.

\FloatBarrier

\section{Substitution-based baselines and controls}
\label{sec:supp_terminal_substitution}

\subsection{Experimental setup and substitution methods}
\label{sec:supp_substitution_methods}

All terminal element-substitution experiments start from the full-sampling results of the CSSE-CSP model.
A nominal fraction $r=0.5$ of the atomic sites, approximately matched to the terminal type-change fraction of CSSE-CSP, is selected uniformly at random, and only the element types at these sites are modified.

\textbf{Codebook-based substitution.}
Replacement probabilities are constructed from the cosine similarities between element codewords in the CSSE-CSP codebook.
For an original element $a$, the probability of sampling a replacement element $b$ is

\begin{equation}
P(b\mid a)=
\frac{\exp(S_{ab}/T_{\mathrm{cb}})}
{\sum_{k\neq a}\exp(S_{ak}/T_{\mathrm{cb}})},
\qquad b\neq a,
\end{equation}

\noindent where $S_{ab}$ denotes the cosine similarity between the codewords of elements $a$ and $b$ and $T_{\mathrm{cb}}=1.0$ is the sampling temperature.

\textbf{Hautier-based substitution.}
Replacement weights are constructed from the Hautier empirical substitution statistics~\cite{hautier2011substitution}, restricted to element pairs that share at least one oxidation state,

\begin{equation}
W_{ab}=
\max_{q\in\mathcal{O}_a\cap\mathcal{O}_b}
P(b^{q}\mid a^{q}),
\end{equation}

\noindent where $\mathcal{O}_a$ and $\mathcal{O}_b$ are the sets of oxidation states of elements $a$ and $b$ in the statistics and $P(b^{q}\mid a^{q})$ is the conditional probability of substituting species $b$ in oxidation state $q$ for species $a$ in the same oxidation state; $W_{ab}=0$ when the two sets do not intersect.
Only the highest-scoring 15\% of the ordered pairs $(a,b)$ with $a\neq b$ and $W_{ab}>0$ are retained, and the replacement element at each selected site is drawn from the same normalised distribution as in the codebook-based scheme, with $S_{ab}$ replaced by $W_{ab}$.
Elements left without an eligible target are not substituted, so the effective replacement fraction is 0.485 rather than the nominal 0.5.

\subsection{Template-based baseline methods}
\label{sec:supp_template_baselines}

The template-based baseline in Table~1 of the main text, data-mined ion exchange, is a non-generative method taken from the \texttt{matgen\_baselines} package; we follow its original implementation and only describe our differences here.

The original package retrieves parent structures on-the-fly from the Materials Project web API.
We instead draw all parent structures from the MP20 training split, so that the reference set matches that of the diffusion models.
For ion exchange, parents are restricted to training structures with an energy above hull of at most $0.1$~eV\,atom$^{-1}$.

The template-based structures are relaxed using MatterSim-1M before evaluation with LeMat-GenBench, following the same relaxation and evaluation pipeline as the diffusion-based results in Table~1 of the main text.
The generation scale and metric definitions likewise follow those used for the diffusion-based results.

\section{Property-guided generation and DFT validation}
\label{sec:supp_property_guidance}

\subsection{Conditional model and classifier-free guidance}
\label{sec:supp_conditional_training}

\textbf{Property conditioning.}
The standardized scalar property $y$ is encoded and added to the diffusion-time embedding as
\begin{equation}
\mathbf{e}_t(y)
=\mathrm{Emb}_{\mathrm{time}}(t)
+\mathrm{MLP}(\mathrm{SinEmb}(y)),
\end{equation}
where $\mathrm{SinEmb}$ is a sinusoidal encoding.
The null condition uses a zero property embedding, and the output layer of the property MLP is initialized to zero.

\textbf{Classifier-free guidance.}
The guided prediction is calculated separately for the lattice, fractional-coordinate and element-type channels as
\begin{equation}
\hat{\mathbf{z}}
=\mathbf{z}(\varnothing)
+w\left[\mathbf{z}(y)-\mathbf{z}(\varnothing)\right],
\qquad
\mathbf{z}\in
\{\boldsymbol{\epsilon}_{L},\,
\mathbf{s}_{F},\,
\boldsymbol{\epsilon}_{T}\},
\label{eq:supp_cfg}
\end{equation}
where $\boldsymbol{\epsilon}_{L}$ and $\boldsymbol{\epsilon}_{T}$ are the predicted lattice and element-type noises, respectively, $\mathbf{s}_{F}$ is the predicted fractional-coordinate score, and $w$ is the guidance scale.

\subsection{Target properties, evaluation and MatterGen comparison}
\label{sec:supp_property_comparison}

For property-guided IILS, the projection time is fixed at $t_s=0.2$ across all three targets.

For comparison, we generate samples using the official pretrained property-conditional MatterGen models and their released sampling code.
All methods follow the same MatterSim-1M relaxation, DFT property evaluation and benchmark criteria, with results given in Supplementary Table~\ref{tab:supp_property_comparison}.

\begin{table*}[t]
\centering
\caption{\textbf{Property-guided generation and comparison with MatterGen.}
MSUN follows the definition used in the main text.
Match denotes the fraction of structures whose evaluated property lies within $\pm 10\%$ of the target value, where the targets are 3.0~eV, 0.20~\AA$^{-3}$ and 250~GPa for band gap, magnetic density and bulk modulus, respectively, and Both denotes the fraction satisfying both criteria.
All percentages are calculated over the structures whose DFT evaluation completed, namely 100 for magnetic density and bulk modulus and, for band gap, 300 for CSSE-CSP with and without IILS, 298 for OneHot and 296 for MatterGen.
}
\label{tab:supp_property_comparison}

\resizebox{\textwidth}{!}{%
\begin{tabular}{lccccccccc}
\toprule
& \multicolumn{3}{c}{Band gap (eV)}
& \multicolumn{3}{c}{Magnetic density (\AA$^{-3}$)}
& \multicolumn{3}{c}{Bulk modulus (GPa)} \\
\cmidrule(lr){2-4}
\cmidrule(lr){5-7}
\cmidrule(lr){8-10}
Model
& MSUN (\%)
& Match (\%)
& Both (\%)
& MSUN (\%)
& Match (\%)
& Both (\%)
& MSUN (\%)
& Match (\%)
& Both (\%) \\
\midrule
OneHot
& 19.8 &  5.0 &  1.0
& 17.0 & 50.0 &  9.0
& 35.0 & 40.0 & 12.0 \\
CSSE-CSP
& 26.0 &  7.7 &  2.0
& 26.0 & 47.0 & 16.0
& 39.0 & 50.0 & 22.0 \\
CSSE-CSP + IILS
& 30.3 &  8.3 &  3.0
& 41.0 & 78.0 & 34.0
& 46.0 & 32.0 & 17.0 \\
MatterGen
& 30.4 &  6.4 &  1.7
& 52.0 & 82.0 & 49.0
& 24.0 & 52.0 & 14.0 \\
\bottomrule
\end{tabular}%
}
\end{table*}

\FloatBarrier

\subsection{DFT validation}
\label{sec:supp_dft_validation}

DFT calculations follow the Materials Project parameter sets described in Methods, with the following property-specific workflows:

\begin{itemize}
    \item \textbf{Structure relaxation and static calculations.}
    Structures are relaxed using \texttt{MPRelaxSet}, followed by static calculations using \texttt{MPStaticSet.from\_prev\_calc}.
    Magnetic density is calculated as $\rho_{\mathrm{mag}}=M/V$, where $M$ is the total magnetization from the spin-polarized calculation and $V$ is the cell volume.

    \item \textbf{Band gaps.}
    Band structures are calculated using \texttt{MPNonSCFSet.from\_prev\_calc} in line mode, starting from the converged static charge density.
    Band gaps are obtained from the resulting PBE band structures.

    \item \textbf{Bulk moduli.}
    Elastic tensors are calculated from 24 deformed structures without symmetry reduction, using normal strains of $\pm0.5\%$ and $\pm1\%$ and shear strains of $\pm3\%$ and $\pm6\%$.
    Bulk moduli are reported as the Voigt--Reuss--Hill averages $K_{\mathrm{VRH}}$.
\end{itemize}
\FloatBarrier

\section{Additional experimental findings}
\label{sec:supp_additional}

\subsection{Combination with an auxiliary representation-alignment objective}
\label{sec:supp_repa}

We further examine whether IILS remains effective when the generative model is trained with an auxiliary representation-alignment objective. 
Specifically, we adopt CrystalREPA~\cite{zhang2026crystalrepa}, which aligns atom-wise hidden representations with frozen universal-MLIP features through an element-aware contrastive objective. 
All other training and sampling settings follow the corresponding DiffCSP configurations used in this study.
Supplementary Table~\ref{tab:supp_repa} reports the corresponding results.

\begin{table}[htbp]
    \centering
    \caption{\textbf{Combination of the projection with an auxiliary representation-alignment objective (DiffCSP, MP20).}
    The training objective of the REPA rows additionally contains the CrystalREPA alignment term~\cite{zhang2026crystalrepa}; all other settings are aligned with the reference rows, which are those of Table~1 of the main text.
    LT denotes longer training, as in Table~1; the two longer-training rows both use the 50,000-epoch checkpoint so that the reference and REPA settings share the same training budget.}
    \label{tab:supp_repa}
    \small
    \resizebox{\textwidth}{!}{%
    \begin{tabular}{lrrrrrrrrrr}
        \toprule
        Model / Setting &
        Valid$\uparrow$ &
        Unique$\uparrow$ &
        Novel$\uparrow$ &
        $E_{\mathrm{form}}\downarrow$ &
        $E_{\mathrm{hull}}\downarrow$ &
        RMSD$\downarrow$ &
        Stable$\uparrow$ &
        SUN$\uparrow$ &
        Metastable$\uparrow$ &
        MSUN$\uparrow$ \\
        \midrule
        \multicolumn{11}{l}{\textbf{Reference}} \\
        \addlinespace[1pt]
        CSSE-CSP                & 97.5 & 95.9 & 51.9 & -0.84 & 0.08 & 0.21 &  4.0 & 0.7 & 67.6 & 24.9 \\
        \quad + IILS            & 98.9 & 95.9 & 77.9 & -0.73 & 0.11 & 0.18 &  1.9 & 1.0 & 61.8 & 42.4 \\
        \quad + IILS (LT)       & 99.4 & 96.8 & 77.6 & -0.76 & 0.08 & 0.13 &  1.6 & 0.9 & 71.6 & 51.4 \\
        \midrule
        \multicolumn{11}{l}{\textbf{With REPA}} \\
        \addlinespace[1pt]
        CSSE-CSP + REPA         & 97.8 & 95.5 & 43.8 & -0.87 & 0.06 & 0.15 &  4.2 & 0.5 & 74.6 & 22.4 \\
        \quad + IILS            & 99.1 & 95.8 & 76.6 & -0.67 & 0.09 & 0.16 &  1.9 & 0.9 & 67.9 & 46.7 \\
        \quad + IILS (LT)       & 99.3 & 96.6 & 77.4 & -0.75 & 0.08 & 0.14 &  2.0 & 1.3 & 72.8 & 52.5 \\
        \bottomrule
    \end{tabular}%
    }
\end{table}

IILS substantially improves MSUN with and without the auxiliary representation-alignment objective (Supplementary Table~\ref{tab:supp_repa}). 
At the standard training budget, combining IILS with REPA further increases MSUN from 42.4\% to 46.7\%, whereas the difference becomes smaller after continued training (51.4\% to 52.5\%).
These results indicate that the benefits of IILS extend to models trained with an additional representation-alignment objective.

\ifdefined\arxivbuild\else
\bibliographystyle{unsrtnat}
\bibliography{../references}
\fi

\end{document}

\else
\section{Author contributions}

Y.J., H.W. and C.Z. conceptualized this project and wrote the paper.
Y.J. developed the code and performed all the research and analysis.

\section{Competing interests}

The authors declare no competing interests.

\section{Additional information}

Correspondence and requests for materials should be addressed to Han Wang.
Supplementary information is available for this paper.
\fi

\bibliographystyle{unsrtnat}
\bibliography{references}

\end{document}